\documentclass[aps,pre,floatfix,showpacs,twocolumn,eqsecnum]{revtex4}
\usepackage{graphicx}
\usepackage{dcolumn}

\usepackage{amsmath,amssymb}
\usepackage{epsfig}

\newcommand{\p}{\partial}
\newcommand{\ud}{\mathrm{d}}
\newcommand{\bm}{\boldsymbol}

\begin{document}

\title{Theory of Light Emission in Sonoluminescence as Thermal Radiation}

\author{Wang-Kong Tse\footnote{Present address: Condensed Matter Theory Center,
Department of Physics, University of Maryland at  College Park,
College Park, Maryland 20742-4111, USA}}
\author{P.T. Leung\footnote{email: ptleung@phy.cuhk.edu.hk}}
\affiliation{Physics Department and Institute of Theoretical
Physics, The Chinese University of Hong Kong, Shatin, Hong Kong
SAR, China}

\begin{abstract}
Based on the model proposed by Hilgenfeldt {\it at al.} [Nature
{\bf 398}, 401 (1999)], we present here a comprehensive theory of
thermal radiation in single-bubble sonoluminescence (SBSL). We
first invoke the generalized Kirchhoff's law to obtain the thermal
emissivity from the absorption cross-section of a multilayered
sphere (MLS). A sonoluminescing bubble, whose internal structure
is determined from hydrodynamic simulations, is then modelled as a
MLS and in turn the thermal radiation is evaluated. Numerical
results obtained from simulations for argon bubbles show that our
theory successfully captures the major features observed in SBSL
experiments.
\end{abstract}
\date{\today}

\maketitle

\section{Introduction}\label{intro}

Single-bubble sonoluminescence (SBSL or simply SL), first
discovered in 1989, is a phenomenon of periodic light emission by
an oscillating gas bubble trapped in the pressure antinode of a
standing ultrasound wave in water (or other fluids) (see
\cite{Barber:1,Lohse:5} for detailed reviews on SBSL). The
oscillating bubble is stable enough to survive many days through
billions of acoustic cycles while the produced flashes are highly
regular and incandescent. The width of the emitted light pulse
 is around $
10\,-\,100\,{\rm ps}$ with a peak power of the order $10\,{\rm
mW}$ \cite{Barber:1,Lohse:5,Barber:2,pulse:4}. The light pulse has
nearly a gaussian shape with a slight asymmetry, which is
basically identical in the red and UV portions of the spectrum
\cite{pulse:4}, and Hiller {\it et al.} \cite{Hiller:3} further
confirmed that the pulse width and the emission time were
independent of wavelength. However, in an interesting twist, Moran
{\it et al.} \cite{pulse:1} demonstrated that the pulse width did
exhibit a mild dependence on wavelength at $3\,^{\rm o}{\rm C}$,
but not at $24\,^{\rm o}{\rm C}$. Besides, the power spectrum of
the emitted light was found to be broadband without any
characteristic line, decreasing from the UV portion towards the
red in a way that bore a resemblance to a blackbody spectrum
\cite{Hiller:1,Hiller:2,Hiller:3}.

SBSL has become an intriguing topic and an arena for
experimentalists and theorists alike since its discovery.
Numerous attempts have been made to study the bubble motion using
classical bubble dynamics and sophisticated computational fluid
mechanics (CFM)
\cite{Gaitan:2,Lofstedt:1,Lofstedt:2,Wu:1,Moss:1,Moss:2,Yuan:1,Cheng:1}.
It is generally believed that the bubble is heated to temperatures
of tens of thousands and shock waves and plasma could be generated
during the contraction of the bubble.  Various proposals have been
put forward to explain the light emission mechanism, including
surface blackbody radiation
\cite{Putterman:1,Putterman:2,Barber:1,Hiller:1}, neutral and ion
bremsstrahlung \cite{Wu:1,Moss:1,Moss:2,xu,Lohse:1,Lohse:6},
collision-induced emission \cite{Frommhold:3,Frommhold:4}, quantum
vacuum radiation \cite{Eberlein}, confined-electron model
\cite{Bernstein}, proton-tunnelling \cite{Willison:1}, and nuclear
fusion \cite{Barber:4,West:1,West:2}. Some of these proposals,
e.g.
\cite{Putterman:1,Putterman:2,Barber:1,Hiller:1,Wu:1,Moss:1,Moss:2,
xu,Lohse:1,Lohse:6,Frommhold:3,Frommhold:4}, attributed light
emission in SBSL  to the high temperature attained in the bubble
and are classified as thermal radiation schemes  in this paper.
While qualitatively reproducing the spectra detected in SBSL, most
of such thermal schemes failed to explain why the pulse width is
wavelength-independent as measured in some experiments
\cite{pulse:4,Hiller:3}. Owing to this major drawback of thermal
schemes, researchers were forced to look into other non-thermal
and more exotic models (see, e.g.
\cite{Eberlein,Bernstein,Willison:1,Barber:4,West:1,West:2}).

To reconcile the success and the drawback of thermal radiation
schemes, Hilgenfeldt, Grossmann and Lohse \cite{Lohse:1,Lohse:6}
took the finite opacity of the bubble into consideration and
obtained a wavelength-independent pulse width for argon bubbles.
The impact of Hilgenfeldt {\it et al.}'s work is huge and, to some
extent, resurrects the thermal radiation scheme. However, it is
worthwhile to note that the approach adopted in
\cite{Lohse:1,Lohse:6} is deemed a simplified version of the
emission mechanism of SBSL  as several physical processes have not
been included in the proposal \cite{Putterman:3}. For example, the
sonoluminescing bubble is modelled as a uniform one and the fluid
dynamics inside the bubble has been neglected from the outset
\cite{Lohse:1,Lohse:6}. Besides, the Kirchhoff's law used in
\cite{Lohse:1,Lohse:6} to evaluate thermal emissivity of the
bubble has completely ignored the wave nature of light.

In addition, existing literature (e.g.
\cite{Lohse:5,Kirilov:1,Hopkins:1,Flannigan:1}) demonstrates there
exists a gap in the theoretical treatment of light emission
mechanism of sonoluminescence, in that blackbody radiation and
thermal bremsstrahlung are often ascribed as separate possible
mechanisms of SBSL. Often, Planck's formula for blackbody radiation
\cite{Kirilov:1,Hopkins:1} and absorption coefficients for thermal
bremsstrahlung in vacuum \cite{Lohse:1,Lohse:6} are applied
separately in these cases, and the question of whether the bubble is
opaque enough to demonstrate blackbody radiation is argued in a
rather hand-waving manner by comparing the photon mean path with an
estimated size of the light-emitting region. In our view, this is
because of the lack of a single theory which can take account of
both mechanisms in a finite-sized environment (i.e. the bubble) in a
unifying manner. In a recent experimental paper \cite{Flannigan:1},
Flannigan \textit{et al}. demonstrated conclusively the existence of
a plasma state inside the bubble, and hence thermal bremsstrahlung
is an inevitable consequence because of the motion of the electrons
and ions. We emphasize that blackbody radiation and thermal
bremsstrahlung are nothing but one single emission mechanism
manifested upon the degree of optical thickness of the bubble; and
to this purpose, in this paper we have developed a coherent theory
unifying both aspects in the context of SBSL. Within the framework
of this theory, when the bubble becomes optically thick enough, the
thermal bremsstrahlung manifest itself asymptotically as blackbody
radiation.

First, we will consider thermal emission in a realistic
sonoluminescing bubble that is non-uniform in temperature as well as
density. To properly describe processes of thermal
emission and absorption in a finite volume with a size comparable
to the wavelength of light in a consistent manner, the generalized
Kirchhoff's law is used in our paper \cite{Rytov:1}. Second, state-
of-the-art CFM is applied here to evaluate the temperature and
density distributions in the bubble \cite{Yuan:1,Cheng:1}. Through
such elaboration of the thermal radiation scheme
\cite{Lohse:1,Lohse:6}, we succeed in obtaining an emission
spectrum that agrees nicely with the experimental data as
summarized in \cite{Barber:1} and, in addition, resolve the
dilemma of whether the pulse width is dependent on the wavelength.

The rest of this paper is organized as follows. We first present
the generalized Kirchhoff's law in Sec.~\ref{thr}, and show in
Sec.~\ref{GOM} that it leads to the formula for emissivity used in
\cite{Lohse:1,Lohse:6} in certain limits. In Secs.~\ref{MLS} and
\ref{WOM} we make use of the generalized Kirchhoff's law to derive
the spectral radiance of a heated multilayered sphere. In
Secs.~\ref{lem} and \ref{CFM} respectively we furnish the plasma
model and CFM used in the present paper. We then present relevant
numerical results in Sec.~\ref{UPS} and conclude our paper in
Sec.~\ref{sum}.

\section{Thermal Radiation}\label{thr}
In SL, the high temperature reached inside the bubble ionizes the
gas content, forming a partially ionized plasma
\cite{Moss:2,Lohse:1,Lohse:6,Ho:1,Chen:1}. Besides having finite
optical thickness, the bubble has a size $\sim
0.1-1\;\mu\textrm{m}$ near the instant of maximum compression,
which is comparable to the wavelength of the emitted light ranging
from $200\,\textrm{nm}$ to $800\;\textrm{nm}$. Hence, the
assumption of geometric optics is invalid. To properly take
account of the effects of finite absorption, wave reflection and
diffraction, we employ in this paper the generalized
Kirchhoff's law \cite{Rytov:1}, which yields the `classical' form
of Kirchhoff's law as an asymptotic limit, to compute the power
spectrum.

We first review the statement of the `classical' form of
Kirchhoff's law (see, e.g. \cite{Bekefi:1}). Consider an isotropic
absorbing medium with complex dielectric constant $\epsilon =
\epsilon_R+i\epsilon_I$, refractive index $n = \sqrt{\epsilon} =
n_R+in_I$, and a typical size $d$. If the medium is in thermal
equilibrium at temperature $T$ and $d$ is large compared with the
wavelength of light $\lambda$ so that the geometric optics
approximation holds, then the classical form of Kirchhoff's law
relates the emission coefficient $\eta(\omega)$ and the absorption
coefficient $\kappa(\omega) \equiv 2n_I\omega/c$ at (angular)
frequency $\omega \equiv 2\pi f$ as \cite{Bekefi:1}:
\begin{equation}\label{}
{\eta}/{\kappa} = n_R^2B_{\omega}(T),
\end{equation}
where
\begin{equation}\label{}
B_{\omega}(T)
=\frac{\hbar\omega^3}{8\pi^3c^2({e^{\hbar\omega/k_BT}-1})}
\end{equation}
is the (frequency) spectral light intensity of a blackbody for one
polarization, with $c$,  $\hbar$ and $k_B$ being  the speed of
light in vacuum,  the Planck constant $h$ divided by $2\pi$, and
the Boltzmann constant, respectively. We remark that, besides the
assumption $\lambda \ll d$, its application
 is justified only to a volume emitter which is optically thin.

The generalized Kirchhoff's law \cite{Rytov:1} is a generalization
of the classical form of Kirchhoff's law to all sizes $d$ and
optical thickness of a finite-size emitter, and can be derived
from the Maxwell equations:
\begin{eqnarray}
&&\nabla \cdot {\boldsymbol{D}}({\boldsymbol{r}},t) =
\rho_n({\boldsymbol{r}},t), \label{thr2} \\
&&\nabla \cdot {\boldsymbol{B}}({\boldsymbol{r}},t) = 0, \label{thr3} \\
&&\nabla \times
{\boldsymbol{E}}({\boldsymbol{r}},t)+\frac{\partial
{\boldsymbol{B}}({\boldsymbol{r}},t)}{\partial t} = 0, \label{thr4} \\
&&\nabla \times
{\boldsymbol{H}}({\boldsymbol{r}},t)-\frac{\partial
{\boldsymbol{D}}({\boldsymbol{r}},t)}{\partial t} =
{\boldsymbol{J}}_n({\boldsymbol{r}},t). \label{thr5}
\end{eqnarray}
Here the electric field $\bm{E}$, the magnetic induction $\bm{B}$,
the displacement field $\boldsymbol{D}$ and the magnetic field
strength $\boldsymbol{H}$ are generated by the fluctuating charge
density $\rho_n$ and current density $\boldsymbol{J}_n$, which are
direct consequence of the random thermal motions of the charges
and satisfy the fluctuation-dissipation theorem (see, e.g.
\cite{Lifshitz:1}):
\begin{eqnarray}
&&\langle{\tilde{J}_{n,i}({\boldsymbol{r}},\omega)
\tilde{J}_{n,j}^*({\boldsymbol{r}}',\omega)}\rangle \\ \nonumber
&=&
\frac{4\omega}{\pi}\theta(\omega,T)\epsilon_0\epsilon_I({\boldsymbol{r}},\omega)
\delta({\boldsymbol{r}}-{\boldsymbol{r}}')\delta_{ij}.
\label{thr13}
\end{eqnarray}
Here $\tilde{J}_{n,i}({\boldsymbol{r}},\omega)$ is the
$i$-component of the Fourier transform of
${\boldsymbol{J}}_n({\boldsymbol{r}},t)$,
\begin{equation}
\theta(\omega,T)  =
\frac{\hbar\omega}{2}\coth\frac{\hbar\omega}{2k_BT} \label{thr14}
\end{equation}
is the average energy of a quantum harmonic oscillator at
temperature $T$, and $\epsilon_0$ is permittivity constant of
vacuum.

In the presence of the thermal fluctuations $\rho_n$ and
$\boldsymbol{J}_n$, all fields $\boldsymbol{E}$, $\boldsymbol{D}$,
$\boldsymbol{H}$ and $\boldsymbol{B}$ are thermally fluctuating
quantities. The set of stochastic equations,
(\ref{thr2})-(\ref{thr5}), together with the
fluctuation-dissipation theorem, completely determine the
statistics of the electromagnetic field of such a system and the
relevant theory is termed the thermal electromagnetic theory (TET)
\cite{Rytov:1}.

The generalized Kirchhoff's law derived from TET \cite{Rytov:1}
states that the spectral radiance $P_{\lambda}$, defined as the
power emitted per unit wavelength interval and unit solid angle,
is given by the following formula \cite{Rytov:1}:
\begin{equation}
P_{\lambda}({\bf n}) = B_{\lambda}(T)\,\sigma_{\rm abs}({\bf n}).
\label{thr33}
\end{equation}
Here $P_{\lambda}({\bf n})$ is the spectral radiance in the
direction of ${\bf n} \equiv {\bf r}/|{\bf r}|$,  $\sigma_{\rm
abs}({\bf n})$ is the absorption cross-sectional area of the
emitter for an electromagnetic wave illuminating the emitter from
the direction of ${\bf n}$, and
\begin{equation}\label{}
B_{\lambda}(T) = \frac{hc^2}{\lambda^5({e^{{hc}/{\lambda
k_BT}}-1})}
\end{equation}
is the spectral light intensity in each polarization for an ideal
blackbody. In (\ref{thr33}) it is understood that $\sigma_{\rm
abs}$ is the sum of the absorption cross-sectional areas for
incident light waves with two perpendicular polarizations. It is
evident that $\sigma_{\rm abs}$ is equal to the power $P_{a}$
dissipated in the emitter for an incident plane wave carrying unit
energy flux, which can be obtained from the following volume
integral over the emitter:
\begin{equation}
P_{a} = \frac{\omega}{2}\int {\ud}^3r\; \epsilon_0
\epsilon_I({\boldsymbol{r}},\omega)
|{\boldsymbol{E}}_{}({\boldsymbol{r}},\omega)|^2.
\label{thr24}
\end{equation}
It is worthwhile to note that
${\boldsymbol{E}}_{}({\boldsymbol{r}},\omega)$ the electric field
developed inside the emitter might be strongly enhanced at certain
frequencies due to resonance effects and thus carries non-trivial
frequency dependence \cite{Chen:1}.

Besides the spectral radiance, the integrated power of the emitted
light pulse is another quantity measured in SBSL experiments and
is simply the integral of Eq.~(\ref{thr33}) over the wavelength,
\begin{equation}
P = \int_{\lambda_1}^{\lambda_2}{\ud}\lambda\; P_{\lambda}.
\label{thr35}
\end{equation}
\section{Geometric optics model} \label{GOM}
We now apply the generalized
Kirchhoff's law summarized above to consider light emission in the
uniform bubble model (UBM) as proposed in \cite{Lohse:1,Lohse:6}.
We will see that the formula for the spectral radiance used in
\cite{Lohse:1,Lohse:6} is only an approximate one that is valid
only under certain restrictions.

To use a simple situation to elucidate the generalized Kirchhoff's
law, we first calculate the power emitted from a slab. Consider a
weakly absorbing slab illuminated normally by a plane-polarized
plane wave with unit flux. Under the assumption that the slab
(with thickness $L$ and area $A$) has dimensions sufficiently
large so that multiple internal reflections can be neglected and
the internal electric field can be represented by a single
decaying, travelling wave only, the internal power loss can be
found from Eq.~(\ref{thr24}):
\begin{equation}
P_{a} = \frac{{\omega}\;\epsilon_0\epsilon_I}{\kappa c}(1-p_{\rm
ref})\; A {(1-e^{-\kappa L})}, \label{thr36}
\end{equation}
Here $p_{\rm ref}$ is the fraction of power reflected from the
slab surface. For small absorption, the imaginary part of the
refractive index $n_I \simeq \epsilon_I/2$, so
\begin{equation}
P_{\lambda} = 2(1-p_{\rm ref})B_{\lambda}(T)(1-e^{-\kappa L})A,
\label{thr38}
\end{equation}
where the factor of 2 properly takes care of the two possible
polarizations. This result is consistent with that obtained from
the standard radiative transfer theory \cite{Zeld:1,Bekefi:1},
upon which the spectral radiance obtained in
\cite{Lohse:1,Lohse:6} is founded.

In the following we explicitly derive the formula for the spectral
radiance that was used by Hilgenfeldt {\it et al.} in
\cite{Lohse:1,Lohse:6} from Eq.~(\ref{thr38}).
\begin{figure}
\includegraphics[width=5.6cm,angle=0]{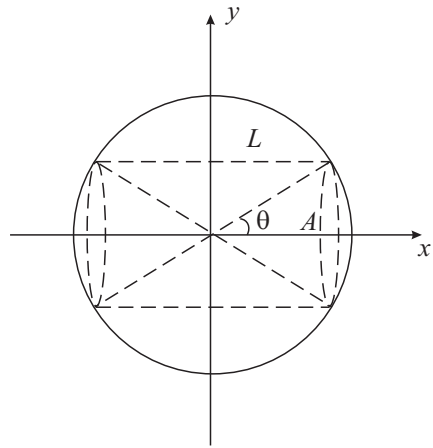}
\caption{\small The configuration of a uniform absorbing sphere.}
\label{thrf1}
\end{figure}
Fig.~\ref{thrf1} shows a uniform sphere of radius $R$, which is
divided into multiple thin cylindrical shells. Each of these
shells, indicated by the dashed lines, subtends an angle of
$\pi-2\theta$ at the center, and has a length $L(\theta) =
2R\cos\theta$ along the $x$-direction and a differential
cross-sectional area $dA = 2\pi R^2\sin\theta \cos\theta d
\theta$. With the assumptions that (i) $p_{\rm ref} = 0$, (ii)
effects of refraction and diffraction at the spherical interface
are negligible, (iii) internal reflection is ignorable, and (iv)
each of these shells can be considered as a slab with area $dA$
and thickness $L(\theta)$, the radiance per unit solid angle
follows directly from Eq.~(\ref{thr38}) is given by:
\begin{eqnarray}
P_{\lambda} &=& \int {\ud}A\; 2B_{\lambda}(T)(1-e^{-\kappa z})
\nonumber \\
&=& 2\pi R^2
B_{\lambda}(T)\left[1+\frac{e^{-2{\kappa}R}}{{\kappa}R}+\frac{e^{-2{\kappa}R}-1}{2{\kappa^2}R^2}\right].
\label{thr40}
\end{eqnarray}
Multiplying this by the total solid angle $4\pi$ straightforwardly
yields the formula Eq.~(17) in Ref.~\cite{Lohse:1}. As seen from
the derivation here, this formula is only valid under the
assumptions mentioned above, through which the wave nature of
light has been completely neglected. Such an emission model will
be referred to as the geometric optics model (GOM) in the
following discussion. As we will show in our numerical results
(Sec.~\ref{UPS}), the conditions for the validity of GOM do not
generally hold in a realistic SL model.

\section{Multilayered Sphere}\label{MLS}
\begin{figure}
\includegraphics[width=5cm,angle=0]{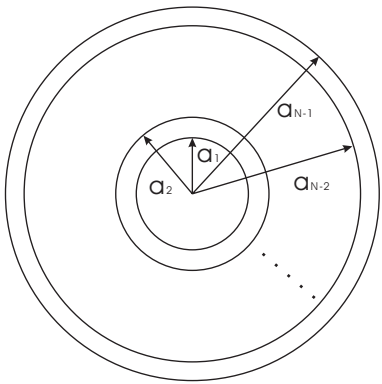}
\caption{\small An $N$-layered spherical system. The inner $N-1$
layers (the bubble) are absorbing, while the outermost layer (the
surrounding fluid) is transparent.} \label{scrf1}
\end{figure}

As discussed in Sect.~\ref{GOM},  Hilgenfeldt {\it et al.}
\cite{Lohse:1,Lohse:6} have assumed UBM as well as GOM in
deriving the spectral radiance. This simplification serves as an
illustration of the essential ingredients in SBSL. Yet its
validity has to be verified, and for realistic calculations
comparable to the experiment, it is necessary to take full account
of the hydrodynamics inside the bubble.  In light of this, in the
following we will employ CFM developed by Ho {\it et al.}
\cite{Ho:1} to simulate the hydrodynamics of the bubble. It is
then obvious that UBM breaks down in such situation and, instead,
we can model the inhomogeneous SL bubble as a multilayered sphere
(MLS) with a piecewise-constant configuration of temperature and
refractive index (see Fig.~\ref{scrf1}). In realistic
calculations, the number of layers is so large that MLS is able to
mimic the continuous distribution obtained from CFM. On the other
hand, MLS also includes UBM as a special case where there is only
a single layer.

Consider an MLS (i.e. the bubble) having a piecewise-constant
temperature profile $T = T_j$, $j = 1,2,{}\dots{},N-1$. As the
absorption cross-section simply becomes
  the sum of the contribution from each  layer,
it follows directly from Eq.~(\ref{thr33}) that:
\begin{equation}
P_{\lambda} = \sum_{j=1}^{N-1} B_{\lambda}(T_j)\,\sigma_{{\rm
abs},j}, \label{thr43}
\end{equation}
where $\sigma_{{\rm abs},j}$ is the contribution of the $j^{\rm
th}$ layer to the total absorption cross-section of the MLS.
Eqs.~(\ref{thr40}) and (\ref{thr43}) are the equations that we use
for computing the power emitted from the SL bubble. In
Sect.~\ref{WOM} we will evaluate $\sigma_{{\rm abs},j}$ from a
wave optics perspective.
\section{Wave optics model} \label{WOM}
To evaluate $\sigma_{{\rm abs},j}$ in an MLS, we first determine the
electromagnetic field inside the absorbing sphere using the
transfer matrix formalism applicable to multilayered configuration
(see, e.g. \cite{Chen:1}).

Consider a circularly polarized plane wave illuminating an
absorbing MLS (see Fig.~\ref{scrf1}). The MLS is composed of $N-1$
spherical shells, and the refractive index, the inner and outer
radii of the $j$-th shell ($j=1,2,\ldots,N-1$) are $n_j$,
$a_{j-1}$ and $a_{j}$, respectively ($a_{0}=0$ and $a_{N-1}=R$ are
assumed). The extended medium surrounding the MLS (i.e. the
bubble) is considered as the $N$-th layer and has a refractive
index $n_N$.

The electric and magnetic fields of the incident wave are given by
${\boldsymbol{E}}_{\rm inc} ({\boldsymbol{r}}) = ({\boldsymbol
{\hat{x}}}+i{\boldsymbol{\hat{y}}})\;\mathrm{exp}({ikz})$ and
 ${\boldsymbol{B}}_{\rm inc}({\boldsymbol{r}})
=-{i}\boldsymbol{E}({\boldsymbol{r}})/c$.
Hereafter we assume that the wave has positive helicity. The
electric and magnetic field inside the $j$-th layer of the MLS (
$j = 1,2,{}\dots{},N$), can be respectively expressed in multipole
expansion as:
\begin{eqnarray}
{\boldsymbol E}_j({\boldsymbol r})
&=&\sum_{l=1}^{\infty}i^l\sqrt{4\pi(2l+1)}
\nonumber \\
&&\times\left\{f_j^{({\rm E})}{\boldsymbol Y}_{l,1}^{(0)}+
\frac{\nabla \times [f_j^{({\rm M})}{\boldsymbol
Y}_{l,1}^{(0)}]}{n_jk}\right\}, \label{scr5} \\
 {\boldsymbol B}_j({\boldsymbol r})
&=&-\frac{in_j}{c}\sum_{l=1}^{\infty}i^l\sqrt{4\pi(2l+1)}
\nonumber \\
&&\times\left\{f_j^{({\rm M})}{\boldsymbol Y}_{l,1}^{(0)}+
\frac{\nabla \times [f_j^{({\rm E})}{\boldsymbol
Y}_{l,1}^{(0)}]}{n_j k}\right\}, \label{scr6}
\end{eqnarray}
where  the vector spherical harmonics
$\boldsymbol{Y}_{lm}^{(0)}(\theta,\phi) \equiv
-{i}\boldsymbol{r}{\times}{\nabla}Y_{lm}(\theta,\phi)/\sqrt{l(l+1)}$,
with $Y_{lm}(\theta,\phi)$ being the ordinary spherical harmonics
\cite{AM:1}.

For the extended medium where $j = N$, $f_N^{({\rm E})} =
f_N^{({\rm M})} = j_l(n_N x)$, with $j_l$ being the spherical
Bessel function of the $l$-th order and $x=kr$,
Eqs.~(\ref{scr5})-(\ref{scr6}) reduce to the ordinary multipole
expansion of a plane wave. For $j < N$, $f_j^{(\mu)}(r)$ ($\mu =
\rm{E}, \rm{M}$) are governed by the radial wave equation:
\begin{eqnarray}
&&\left[\frac{{\ud}^2}{{\ud}r^2}+n_j^2k^2-\frac{l(l+1)}{r^2}\right]rf_j^{(\mu)}(r)
= 0, \label{scr9}
\end{eqnarray}
and, in addition, they satisfy the standard boundary conditions on
electromagnetic fields imposed at $r=a_j$:
\begin{eqnarray}
&&f_j^{({\rm E})}(r) = f_{j+1}^{({\rm E})}(r), \label{scr10} \\
&&\frac{{\ud}}{{\ud}r}\left[rf_j^{({\rm E})}\right] =
\frac{{\ud}}{{\ud}r}\left[rf_{j+1}^{({\rm E})}(r)\right],
\label{scr11}
\\
&&n_jf_j^{({\rm M})}(r) = n_{j+1}f_{j+1}^{({\rm M})}(r), \label{scr13} \\
&&\frac{n_{j+1}}{n_j}\frac{{\ud}}{{\ud}r}\left[rf_j^{({\rm
M})}\right] = \frac{{\ud}}{{\ud}r}\left[rf_{j+1}^{({\rm
M})}(r)\right]. \label{scr14}
\end{eqnarray}

In  general $f_j^{(\mu)}(r)$ can be written as:
\begin{equation}
f_j^{(\mu)}(r) =
\alpha_j^{(\mu)}j_l(n_jkr)+\beta_j^{(\mu)}h_l^{(1)}(n_jkr),
\label{scr15}
\end{equation}
where  $h_l^{(1)}$ is the $l$-th order spherical Hankel function
of the first kind. Substituting Eq.~(\ref{scr15}) into
Eqs.~(\ref{scr10})-(\ref{scr14}) and recasting them into matrix
form, we can show that:
\begin{equation}
\left[
\begin{array}{c}
\alpha_{j+1}^{(\mu)} \\
\beta_{j+1}^{(\mu)}
\end{array}
\right] = {\bf{T}}_j^{(\mu)} \left[
\begin{array}{c}
\alpha_j^{(\mu)} \\
\beta_j^{(\mu)}
\end{array}
\right], \label{scr18}
\end{equation}
where ${\bf{T}}_j^{(\rm{E})}$ and ${\bf{T}}_j^{(\rm{M})}$ are
respectively the TE-mode and TM-mode transfer matrix at $r=a_j$,
given explicitly by:
\begin{widetext}
\begin{eqnarray}
&&{\bf{T}}_j^{(\rm{E})} = -in_{j+1}x_j^2 \times \nonumber \\
&&\left[
\begin{array}{cc}
\;\;W_{j+1,j}^{(\rm{E})}[j_l(n_jx_j),h_l^{(1)}(n_{j+1}x_j)] &
\;\;W_{j+1,j}^{(\rm{E})}[h_l^{(1)}(n_jx_j),h_l^{(1)}(n_{j+1}x_j)]    \\
-W_{j+1,j}^{(\rm{E})}[j_l(n_jx_j),j_l(n_{j+1}x_j)] &
-W_{j+1,j}^{(\rm{E})}[h_l^{(1)}(n_jx_j),j_l(n_{j+1}x_j)]
\end{array}
\right], \label{scr19}
\end{eqnarray}
\begin{eqnarray}
&&{\bf{T}}_j^{(\rm{M})} = -in_{j+1}^2n_jx_j^2 \times
\nonumber \\
&&\left[
\begin{array}{cc}
\;\;W_{j+1,j}^{(\rm{M})}[j_l(n_jx_j),h_l^{(1)}(n_{j+1}x_j)]  &
\;\;W_{j+1,j}^{(\rm{M})}[h_l^{(1)}(n_jx_j),h_l^{(1)}(n_{j+1}x_j)]       \\
-W_{j+1,j}^{(\rm{M})}[j_l(n_jx_j),j_l(n_{j+1}x_j)]        &
-W_{j+1,j}^{(\rm{M})}[h_l^{(1)}(n_jx_j),j_l(n_{j+1}x_j)]
\end{array}
\right]. \label{scr21}
\end{eqnarray}
\end{widetext}
Here $x_j=ka_j$, and for convenience we define the generalized
Wronskian $W_{j+1,j}^{(\mu)}[f,g]$ for TE and TM modes  as:
\begin{eqnarray}
W_{j+1,j}^{(\rm{E})}[f,g] &=& fg'-f'g \label{scr16}, \\
W_{j+1,j}^{(\rm{M})}[f,g] &=&
\frac{fg'}{n_{j+1}^2}-\frac{f'g}{n_j^2}\nonumber
\\&&+(\frac{1}{n_{j+1}^2}-\frac{1}{n_j^2})\frac{fg}{x},
\label{scr17}
\end{eqnarray}
where $' = {{\ud}}/{{\ud}x}$.

In short, the transfer matrix ${\bf{T}}_j^{(\mu)}$ can be written
as:
\begin{equation}
{\bf{T}}_j^{(\mu)} = \left[
\begin{array}{cc}
A_j^{(\mu )} &  B_j^{(\mu )}       \\
C_j^{(\mu )} &  D_j^{(\mu )}
\end{array}
\right]. \label{scr23}
\end{equation}
We can now solve for the field coefficients $\alpha_{j}^{(\mu )}$
and $\beta_{j}^{(\mu )}$using the regularity condition at $r = 0$
and radiation boundary conditions at $r = \infty$, leading to the
following relation:
\begin{equation}
\left[
\begin{array}{c}
1 \\
\beta_{N}^{(\mu )}
\end{array}
\right] = \left[
\begin{array}{cc}
A^{(\mu )} &  B^{(\mu )}       \\
C^{(\mu )} &  D^{(\mu )}
\end{array}
\right] \left[
\begin{array}{c}
\alpha_1^{(\mu )} \\
0
\end{array}
\right], \label{scr24}
\end{equation}
Here $A^{(\mu )}, B^{(\mu )}, C^{(\mu )}, D^{(\mu )}$ (without the
subscript $j$) are the elements of the \textit{total} transfer
matrix ${\bf{T}}^{(\mu)}$ from the layer $j = 1$ to the layer $j =
N-1$, i.e. ${\bf{T}}^{(\mu)} =
{\bf{T}}_{N-1}^{(\mu)}{\bf{T}}_{N-2}^{(\mu)}{}\cdots{}{\bf{T}}_{1}^{(\mu)}$.
From Eq.~(\ref{scr24}) we immediately get
\begin{eqnarray}
\alpha _1^{(\mu )}  &=& (A^{(\mu )})^{-1},
\label{scr26} \\
\beta _N^{(\mu )}  &=& C^{(\mu )}\alpha _1^{(\mu )} = C^{(\mu
)}/A^{(\mu )}. \label{scr27}
\end{eqnarray}
The field coefficients of each layer $\alpha_{j}^{(\mu )}$,
$\beta_{j}^{(\mu )}$ is now readily obtained
 by iteratively applying Eq.~(\ref{scr18}), from $ j =
1$ to $j = N-1$, using the boundary conditions
Eqs.~(\ref{scr26})-(\ref{scr27}).

Now we proceed to calculate the absorption cross-section of the
$j$-{\rm th} spherical shell. To this end, we first evaluate:
\begin{equation}
{\cal F}_{ j} = \frac{n_N ca_{j}^2}{2}{\rm Re}\left\{\int
{\ud}\Omega
\boldsymbol{E}_j\cdot(\boldsymbol{\hat{r}}\times\boldsymbol{B}_j^*)\right\},
\label{scr28}
\end{equation}
which is directly proportional to the energy flux ${\cal F}_{ j}$
crossing the $j$-th interface at $r=a_j$.
 With Eqs.~(\ref{scr5}),
(\ref{scr6}), we find, after some algebraic manipulations,
\begin{eqnarray}
{\cal F}_{ j} &=& \frac{2\pi x n_N}{k^2} {\rm
Im}\left\{n_j^*\sum_{l=1}^{\infty}(2l+1) \times\right.
\nonumber \\
&&\left.\left\{\frac{f_j^{(E)}[xf_j^{(E)}]'^*}{n_j^*}
-\frac{f_j^{(M)*}[xf_j^{(M)}]'}{n_j}\right\}\right\},
\label{scr29}
\end{eqnarray}
where $x$ is evaluated at $r=a_j$. Following directly from energy
conservation, the absorption cross-section of the $j$-th spherical
shell, $\sigma_{{\rm abs},j}$ is given by the difference between
${\cal F}_{j}$ and ${\cal F}_{j-1}$:
\begin{equation}
\sigma_{{\rm abs},j} = {\cal F}_{j}-{\cal F}_{j-1}. \label{scr30}
\end{equation}

Eqs.~(\ref{scr29})-(\ref{scr30}) in conjunction with
Eqs.~(\ref{thr35}) and (\ref{thr43}), with no simplifying
assumptions based on geometric optics and optical thickness, are
the major results of the wave optics model (WOM) introduced here,
which will be used in this paper to calculate the light emission
of a SL bubble.
\section{Plasma Model of Multilayered Sphere} \label{lem}
\subsection{Formation of Plasma}
In order to consider light emission processes in an MLS, we have to
find the complex refractive index $n_j$ and hence the absorption
coefficient $\kappa_j$ of each layer. In the presence of the high
temperature developed during the collapse of a SL bubble, the gas
inside the bubble is partially ionized at the instant of light
emission \cite{Moss:1,xu,Lohse:1,Lohse:6,Ho:1,Chen:1}. Here we
adopt a simple collision-dominated plasma model in which the
collision frequency $\nu$ is a constant dependent on the
concentration and temperature of the plasma \cite{Bekefi:1}. The
refractive index is then given by:
\begin{equation}
n^2(\omega) = n_{b}^2-\frac{\omega_p^2}{\omega^2+i\nu\omega},
\label{lem1}
\end{equation}
where $\omega_p = \sqrt{{{N_e}e^2}/{m\epsilon_0}}$ is the plasma
frequency,
\begin{equation}
n_{b} =
\left(\frac{1+{2N_0\alpha}/{3\epsilon_0}}{1-{N_0\alpha}/{3\epsilon_0}}\right)^{\frac{1}{2}},
\label{lem3}
\end{equation}
is the contribution to the refractive index due to the background
neutral atoms as given by the Clausius-Mossotti equation (see,
e.g. \cite{Jackson:1}), and $\alpha$ is the atomic polarizability,
with $e$ and $m$ being the charge and the mass of electron,
respectively. For convenience we have suppressed the
$j$-dependence of the relevant physical quantities in the above
equations. In the subsequent discussion we will use CFM developed
in \cite{Ho:1} to determine the number densities of electron and
atom, $N_e$ and $N_0$, as a function of time.
\subsection{Collision Processes in plasma}\label{PS}
Around the instant of maximum compression, the temperature of a
sonoluminescing bubble could reach tens of thousands of kelvins,
ionizing the gas content inside. While Ref.~\cite{Lohse:1} showed
that the fraction of Ar$^{+}$ ions amounts to less than $1\%$ of
the bubble content using a uniform bubble assumption;
Ref.~\cite{Ho:1} took account of full hydrodynamics and showed
that this fraction can be as large as $30\%$. However, both papers
point to the fact that the bubble content becomes a partially
ionized plasma with Ar being the dominant species. Under this
condition, bremsstrahlung is thought to play a major role in the
light emission mechanism; in particular, electron-atom
bremsstrahlung is expected to be the dominant process compared to
other bremsstrahlung processes \cite{Lohse:1, Lohse:6}.

On the other hand, as the numerical results obtained from CFM
showed the degree of ionization may be much higher than those from
UBM, we have developed here the WOM to take account of effects due
to reflection, refraction and diffraction as well. In accordance
with the approach of WOM, one has to consider the total effective
collision frequency $\nu$, which is the sum of contributions from
electron-ion collision, electron-atom collision and electron-ion
recombination.  According to the initial and final states of the
electron, the first two are also known as free-free transitions,
the latter one as free-bound transition. Of course, the direct
product of such collisions is the emission of photons and the
corresponding mechanisms are  electron-ion bremsstrahlung,
electron-atom bremsstrahlung as well as electron-ion
recombination. Therefore, one can easily see the difference, as well as
the relation between our approach and that proposed in \cite{Lohse:1,
Lohse:6}. In the following we provide the formulas of the
collision frequencies in these processes.

\subsubsection{Electron-ion collision}
The simplest picture describing electron-ion collisions and
electron-ion bremsstrahlung is to regard them as
individual binary events so that collective phenomena do not
enter. Under such an assumption the differential emission
cross-section $d\sigma_f$, which measures the probability of light
emission due to the scattering of a unit incident electron flux
from an ion with charge $Ze$, is given by the well-known Kramer's
formula \cite{Zeld:1}. It is customary to include quantum
mechanical corrections to this classical formula as the Gaunt
factor \cite{Bekefi:1}. For free-free transitions, including the
free-free Gaunt factor $g_{\rm ff}(f,v)$ gives $d\sigma_f$ as:
\begin{equation}
\frac{{\ud}\sigma_f}{{\ud}f} = \left(\frac{e^2}{4\pi\epsilon_0
\hbar c}\right)^3 \frac{16\pi
\hbar^2}{3\sqrt{3}m^2v^2}\frac{Z^2}{f} g_{\rm ff}(f,v),
\label{ps3}
\end{equation}
where $f$ is the light frequency and $v$ is the speed of the
incident electron. Within the range of optical frequencies, the
free-free Gaunt factor is usually of the order 1.

The collisions between charged particles is formally described by
the scattering cross-section $\sigma(\theta)$, measuring the
probability of scattering at an angle $\theta$ of a unit incident
electron flux from an ion. However, the cross-section used in
transport theory to predict the scattering frequency is
$\sigma_{\rm tr}$, related to the former through the relation
$\sigma_{\rm tr} =
\overline{\sigma(\theta)}(1-\overline{\cos\theta})$, where the
over-bar indicates averaging over the scattering angle $\theta$.
Using the relation between the differential emission cross-section
${\ud}\sigma_f$ and the transport cross-section $\sigma_{\rm tr}$
\cite{Zeld:1}
\begin{equation}
\frac{{\ud}\sigma_f}{{\ud}f} =
\frac{8}{3}\frac{e^2v^2}{4\pi\epsilon_0 c^3hf}\sigma_{\rm tr},
\label{ps4}
\end{equation}
the transport cross-section is found to be
\begin{equation}
\sigma_{\rm tr} =
\frac{4{\pi}^2}{\sqrt{3}}\left(\frac{Ze^2}{4\pi\epsilon_0
mv^2}\right)^2 g_{\rm ff}(f,v). \label{ps5}
\end{equation}
The collision frequency, defined as $\nu = Nv\sigma_{\rm tr}$,
where $N$ is the number density of the background species ($N_i$
for ions or $N_0$ for atoms), is then
\begin{equation}
\nu_{\rm ei} =
\frac{4{\pi}^2}{\sqrt{3}}\frac{N_iZ^2e^4}{(4\pi\epsilon_0)^2
m^2v^3} g_{\rm ff}(f,v). \label{ps7}
\end{equation}

\subsubsection{Electron-ion Recombination}
In electron-ion recombination, an electron is first captured by an
ion, forming a bound state with energy levels labelled by quantum
number $n$. A photon is released in such a process and the
differential emission cross-section is given by Eq.~(\ref{ps3})
with the free-free Gaunt factor replaced by the free-bound Gaunt
factor, $g_{\rm fb}(n,f,v)$. In addition to its dependence on the
photon frequency $f$ and the velocity $v$ of the incident
electron, the free-bound Gaunt factor is also a function of $n$
and  approximately of the order unity in optical frequencies.
Accordingly, the transport cross-section and the electron-ion
recombination collision frequency are given respectively by
Eqs.~(\ref{ps5}) and (\ref{ps7}) with $g_{\rm fb}(n,f,v)$
replacing $g_{\rm ff}(f,v)$.
\subsubsection{Electron-atom collision}
An electron moving near a neutral atom can also experience a
short-range Coulomb field, emitting radiation commonly known as
electron-atom bremsstrahlung. There is no simple theory to
determine the corresponding transport cross-section as we are
aware of; and common practice is to determine it from experiment
with different incident electron energies. With good accuracy in
the relevant range of electron energies for an argon SL bubble,
$\sigma_{\rm tr}$ has a linear dependence on the electron energy
$E_e = mv^2/2$ and  $\sigma_{\rm tr} = c_{\rm tr}mv^2/2+d_{\rm
tr}$, with the empirical constants $c_{\rm tr} \simeq 0.1\, {\rm
m^2\,J^{-1}}$ and $d_{\rm tr} \simeq -0.6\times10^{-20}\, {\rm
m^2}$ \cite{Lohse:1,Brown:1}. The electron-atom collision
frequency is therefore
\begin{equation}
\nu_{\rm ea} = N_0v(c_{\rm tr}mv^2/2+d_{\rm tr}). \label{ps12}
\end{equation}
\subsection{Effective collision frequency}
Assuming local thermodynamic equilibrium prevails in the plasma
(as in the case for SL) and weak damping ($\nu \ll \omega$), we
proceed to calculate the effective collision frequency
\cite{Bekefi:1} defined by $\nu^{\rm eff} = \overline{\nu} =
N\overline{v\sigma_{\rm{tr}}(v)}$, here $\overline{\cdots}$
indicates the Maxwellian average:
\begin{eqnarray}
\overline{\cdots}&=& \frac{4\pi}{3}\left(\frac{m}{2\pi
k_BT}\right)^{\frac{3}{2}} \left(\frac{m}{k_BT}\right) \nonumber
\\ &&\times  \int_0^{\infty}{\ud}v\,v^4e^{-{mv^2}/{2k_BT}}(\cdots).
\label{ps14}
\end{eqnarray}
Accordingly, the effective collision frequencies for electron-ion
bremsstrahlung, electron-ion recombination and electron-atom
bremsstrahlung are obtained as (for clarity we drop the
superscript `eff'):
\begin{eqnarray}
\nu_{\rm ei(rc)} &=&
2\left(\frac{2\pi}{3}\right)^{\frac{3}{2}}N_i\left(\frac{Ze^2}{4\pi\epsilon_0
k_BT}\right)^2\left(\frac{k_BT}{m}\right)^{\frac{1}{2}} \nonumber \\
&&\times e^{-hf/k_BT} \overline{g_{\rm ff(fb)}}(f,T), \label{ps15} \\
\nu_{\rm ea} &=& \frac{8\sqrt{2}}{3}N_0\left(\frac{kT}{\pi
m}\right)^{\frac{1}{2}}(3c_{\rm tr}k_BT+d_{\rm tr}). \label{ps17}
\end{eqnarray}
Here the exponential factor $\exp(-hf/k_BT)$ is commonly referred
to as the Cillie exponential factor \cite{Heald:1};
$\overline{g_{\rm ff}}(f,T)$ and $\overline{g_{\rm fb}}(f,T)$ are
the velocity-averaged free-free and free-bound Gaunt factor
respectively \cite{Hulst:1}:
\begin{eqnarray}
\overline{g_{\rm ff}}(f,T) &=& \frac{e^{hf/k_BT}}{k_BT}\int_{
hf}^{\infty}{\ud}E g_{\rm ff}(f,E)e^{-E/k_BT},
\label{ps18} \\
\overline{g_{\rm fb}}(f,T) &=&
2x_1\sum_{n^*}^{\infty}\frac{1}{n^3}e^{x_n}g_{\rm fb}(n,f,v).
\label{ps19}
\end{eqnarray}
In (\ref{ps19}), $x_n = E_{\rm ion}/n^2k_BT$, with $E_{\rm ion}$
being the first ionization energy of the atom, and $n^*$ is the
lowest level for which $E_{n^*} < hf$.

Note that, assuming $g_{\rm ff}$, $g_{\rm fb} \simeq 1$, the above
equations can be reduced to
\begin{eqnarray}
\overline{g_{\rm ff}}(f,T) &=& 1, \label{ps20} \\
\overline{g_{\rm fb}}(f,T) &=&
2x_1\sum_{n^*}^{\infty}\frac{1}{n^3}e^{x_n}. \label{ps21}
\end{eqnarray}
The summation in Eq.~(\ref{ps21}) can be further simplified if the
photon energy is small compared with the ionization energy, so
that the energy level $n^*$ is high in comparison with the ground
state, as is the case for argon \cite{Lohse:1}. Since the density
of the levels increases rapidly with increasing $n$, the discrete
levels higher than $n^*$ can be replaced by a continuous spectrum
and the summation replaced by an integration, and Eq.~(\ref{ps21})
simplifies to
\begin{eqnarray}
\overline{g_{\rm fb}}(f,T) &=& e^{{hf}/{k_BT}}-1 \nonumber \\
&=& e^{{hc}/{k_BT {\rm max}\{\lambda,\lambda_2\}}}-1. \label{ps23}
\end{eqnarray}

As the absorption coefficient $\kappa$ is related to the effective
collision frequency as:
\begin{equation}
\kappa = (\frac{\omega_p}{\omega})^2\frac{\nu}{c}, \label{ps24}
\end{equation}
we can find the absorption coefficients corresponding to
Eqs.~(\ref{ps15})-(\ref{ps17}):
\begin{eqnarray}
\kappa_{\rm ei(rc)} &=&
\frac{4}{3}(\frac{2\pi}{3mk_BT})^{\frac{1}{2}}\frac{Z^2N_i^2e^6\lambda^2}{(4\pi\epsilon_0)^3k_BTc^3m}
\nonumber \\
&&\times e^{-hf/k_BT}\overline{g_{\rm ff(fb)}}(f,T), \label{ps25}
\\ \kappa_{\rm ea} &=& 8\sqrt{2}\frac{N_i
N_0e^2\lambda^2}{4\pi\epsilon_0c^3}(\frac{k_BT}{\pi
m})^{\frac{3}{2}}\nonumber \\ && \times
(c_{tr}+\frac{d_{tr}}{3k_BT}), \label{ps26}
\end{eqnarray}
The absorption coefficient $\kappa_{\rm ei(rc)}$ in
Ref.~\cite{Lohse:1} differs from ours  in two ways. (i) They
differ by a factor of $k_BT/hf$. However, this difference is not
very significant as $k_BT \simeq hf$ in SL; (ii) For electron-ion
bremsstrahlung, the free-free Gaunt factor and the Cillie
exponential correction were neglected in Ref.~\cite{Lohse:1}. As
we will discuss later (see Fig.~\ref{upsf8}), this could result in
a factor of $3$ difference in the calculated spectra.

In the present paper, we explicitly take account of all Gaunt
factors and the exponential correction, in particular we adopt the
fitting formula proposed by Itoh ${\it et~al.}$
\cite{Itoh:1,Itoh:2} to compute the average free-free Gaunt
factor. For the free-bound Gaunt factor, since the photon energies
$1.5-6.2 \, {\rm eV}$, corresponding to the wavelength $200-800
\,{\rm nm}$, is small compared with the ionization energy ($E_{\rm
ion} = 15.8 \, {\rm eV}$ for argon), Eq.~(\ref{ps23}) is still a
good approximation and we retain it for computing the average
free-bound Gaunt factor. As a remark, following Eqs.~(\ref{ps15}),
(\ref{ps20}) and (\ref{ps23}) we have $\nu_{\rm rc}/\nu_{\rm ei}
\simeq e^{{hf}/{k_BT}}-1$, hence electron-ion bremsstrahlung is
more dominant over recombination when $hf \ll k_BT$. For SL,
however, both processes are important since the thermal energy is
typically $\sim 1.7-4.3 \, {\rm eV}$.

For the purpose of comparison, in the following discussion we will use
two different sets of formulas, respectively denoted by P1 and P2
models, to calculate the collision frequencies.  The P1 model
employs the free-bound Gaunt factor, but ignore the  free-free
Gaunt factor and the Cillie exponential cut-off factor:
\begin{eqnarray}
\nu_{\rm ei} &=&
2(\frac{2\pi}{3})^{\frac{3}{2}}N_i(\frac{Ze^2}{4\pi\epsilon_0
k_BT})^2(\frac{k_BT}{m})^{\frac{1}{2}}\frac{k_BT}{hf}  \label{A4}
\\
\nu_{\rm rc} &=&
2(\frac{2\pi}{3})^{\frac{3}{2}}N_i(\frac{Ze^2}{4\pi\epsilon_0
k_BT})^2(\frac{k_BT}{m})^{\frac{1}{2}}\frac{k_BT}{hf} \nonumber \\
&&\times \overline{g_{\rm fb}}(f,T), \label{A5}
\\
\nu_{\rm ea} &=& \frac{8\sqrt{2}}{3}N_a(\frac{k_BT}{\pi
m})^{\frac{1}{2}}(3c_{tr}k_BT+d_{tr}). \label{A6}
\end{eqnarray}
Through (\ref{ps24}), it is obvious that this set of formulas for
the collision frequencies are consistent with the set of
absorption coefficients used in \cite{Lohse:1,Lohse:6}.

By contrast, the P2 model, derived earlier in this section,
readily takes account of free-bound Gaunt factor, free-free Gaunt
factor and exponential cut-off factor:
\begin{eqnarray}
\nu_{\rm ei} &=&
2(\frac{2\pi}{3})^{\frac{3}{2}}N_i(\frac{Ze^2}{4\pi\epsilon_0
k_BT})^2(\frac{k_BT}{m})^{\frac{1}{2}} \nonumber \\
&&\times e^{-hf/k_BT}\overline{g_{\rm ff}}(f,T), \label{A16}
\\
\nu_{\rm rc} &=&
2(\frac{2\pi}{3})^{\frac{3}{2}}N_i(\frac{Ze^2}{4\pi\epsilon_0
k_BT})^2(\frac{k_BT}{m})^{\frac{1}{2}} \nonumber \\
&&\times e^{-hf/k_BT}\overline{g_{\rm fb}}(f,T), \label{A17}
\\
\nu_{\rm ea} &=& \frac{8\sqrt{2}}{3}N_a(\frac{k_BT}{\pi
m})^{\frac{1}{2}}(3c_{tr}k_BT+d_{tr}). \label{A18}
\end{eqnarray}

Finally we state the basic assumptions underpinning the above formulas for
plasma collision processes: (1) the plasma is `cold' meaning that the
electron thermal velocity is negligible with respect to the
phase velocity of the wave, $v_{\rm th} \ll v_{\rm ph}$; (2)
 the plasma is in the weak coupling regime,
i.e. the ions are weakly interacting during their thermal motions.
It is customary to indicate the degree of coupling by the
dimensionless ion-coupling parameter (see, e.g. \cite{Fortov:1}):
\begin{equation}\label{couple}
 \Gamma = \frac{Z^2e^2}{4\pi \epsilon_0
R_{\rm ion}k_BT},
\end{equation}
with $R_{\rm ion} = (4\pi N_i/3)^{-\frac{1}{3}}$ being the mean
inter-ionic distance. If $\Gamma \ll 1$, the system is said to be
weakly coupled; on the other hand if $\Gamma \gtrsim 1$, the
system is in the strong coupling regime \cite{Fortov:1}. We find
$\Gamma \lesssim 1$ in a typical SL bubble (Sec.~\ref{UPS}), hence
the assumption of weak coupling is at least approximately
satisfied.
\section{Computational fluid mechanics}\label{CFM}
In this section, we summarize the CFM used in the present paper,
which was developed by Cheng {\it et al.}~\cite{Yuan:1,Cheng:1}
and later extended by Ho {\it et al.} \cite{Ho:1} to include the
ionization and recombination processes. The model couples the
Rayleigh-Plesset (RP) equation for the bubble wall with the
Navier-Stokes (NS) equations for the gas (including all the
charged species resulting from ionizations), while independently
solving the energy equation for the surrounding water. The number
densities of the charged species are computed from the reaction
rates approach. The effects of viscosity, surface tension,
equation of state (EOS),  compressibility and thermal conductivity
of the ambient liquid are also taken into account.
\subsection{Bubble-wall dynamics}
To account for the effect of liquid compressibility, a more robust
RP equation that gives the bubble radius $R$ as a function of time
$t$ is used \cite{Yuan:1,Cheng:1,Ho:1,kama87}:
\begin{equation}
\frac{1-M}{1+M} R\ddot{R}+\frac{3-M}{2(1+M)}\dot{R}^2 =
H_l-\frac{P_s\left(t'\right)}{\rho_{0}}
  +\frac{t_R\dot{H_l}}{1+M}.
\label{rp3}
\end{equation}
Here  $t_R \equiv R/c_l$, with $c_l$ the speed of sound in the
surrounding liquid, $M \equiv \dot{R}/c_l$, $t' \equiv t+t_R$,
$\rho _0$ is the ambient liquid density, and $P_s(t')=-P_a
\sin(\omega t')$ is the sonic driving pressure with frequency
$\omega$ and amplitude $P_a$. Also, the enthalpy $H_l$ and the
speed of sound of the liquid  and $c_l$ are given by:
\begin{eqnarray}
H_l &= &\int_{P_0}^{P_l}\frac{{\ud}P_l}{\rho_l}, \label{enthalpy1}
\\
{c_l}^2 &=& \frac{{\ud} P_l}{{\ud} \rho_l}. \label{speed1}
\end{eqnarray}
This modified RP equation includes terms to first order in the
Mach number $M$ of the bubble wall and allows for a variable
$c_l$.

Combining Eqs.~(\ref{enthalpy1}), (\ref{speed1}) with the EOS of
the ambient liquid in the modified Tait form \cite{pros86},
\begin{equation}
 \frac{P_l+B}{P_0+B}
=\left(\frac{\rho_l}{\rho_0}\right)^n,  \label{tait}
\end{equation}
yields the explicit forms for $H_l$ and $c_l$:
\begin{align}
H_l &= \frac{n}{n-1}\left(\frac{P_l+B}{\rho_l}
                        -\frac{P_0+B}{\rho_0}\right), \label{enthalpy2}
\\[16pt]
{c_l}^2 &= \frac{n(P_l+B)}{\rho_l}, \label{speed2}
\end{align}
where $B=3049.13$~bar and $n=7.15$ are valid for water up to
$10^5$~bar.

Eqs.~(\ref{rp3}), (\ref{enthalpy2}) and (\ref{speed2}) must be
supplemented by the boundary condition at the bubble wall, namely,
that the pressure $P_l(t)$ on the liquid side of the gas-liquid
interface differs from the pressure $P(R,t)$ on the gas side of
the gas-liquid interface by the effects of surface tension and the
normal component of viscous stresses \cite{pros86},
\begin{equation}
P(R,t)-\tau_{rr}|_{r=R}=P_l(t)+\frac{4\eta_l \dot{R}}{R}
+\frac{2\sigma}{R}.
\end{equation}

\subsection{Hydrodynamics of gas}
The conservation of mass, momentum and energy for the gas flow in
the spherical bubble is described by the compressible NS
equations. They can be rewritten into a conservative form with
source terms as:
\begin{equation}
\frac{\p\rho}{\p t} + \frac{\p}{\p r}(\rho v) = -\frac{2\rho
v}{r},
 \label{NSr1}
\end{equation}
\begin{eqnarray}
&&\frac{\p(\rho v)}{\p t} + \frac{\p}{\p r}\left(\rho v^2 +
P\right) \nonumber \\
&=& -\frac{2\rho v^2}{r} + \frac{1}{r^2}\frac{\p}{\p
r}(r^2\tau_{rr}) + \frac{\tau_{rr}}{r}, \label{NSr2}
\end{eqnarray}
\begin{eqnarray}
&&\frac{\p(\rho E)}{\p t} + \frac{\p}{\p r}(\rho E+P)v \label{NSr3}
\\ &=& -\frac{2(\rho E+P)v}{r}+ \frac{1}{r^2}\frac{\p}{\p r}
\left[r^2\left(v\tau_{rr}+k\frac{\p T}{\p r}\right)\right].
\nonumber
\end{eqnarray}
Here   $r, \ \rho, \ v, \ P, \ T, \ \tau_{rr}, \ k$  $e$ and
$E=e+v^2/2$ are the radial distance from the center of the bubble,
gas density, radial velocity, pressure, temperature, normal
viscous stress, coefficient of thermal conductivity, the internal
energy and total energy per unit mass, respectively.

If, due to ionizations and recombinations, there exists $N_s$
species inside the bubble, then $N_s-1$ mass conservation
equations for these species must be supplemented with
Eqs.~(\ref{NSr1})-(\ref{NSr3}). In Ref.~\cite{Ho:1}, the maximum
ionization level of the gas atom is taken to be 5, making a total
of $N_s = 7$ species inside the bubble. This is more than adequate
for the present temperature range; in fact, Ho {\it et al.}
\cite{Ho:1} have shown that even the second ionization level can
be safely ignored in practice. Note, since the ion densities
change due to ionizations and recombinations, source terms must be
added to the right hand side of the conservation equations.

For convenience, let $f_j$ be the mass fraction of Ar$^{j+}$ (with
$j=0,1,2,3,4,5$) or electrons (with $j=e$), so that
$\sum_{j=e,0}^5 f_j = 1$. Therefore $\rho f_j$ represents the mass
density of an individual species. The number density of an
individual species is related to its mass fraction by $n_j=\rho
f_j/m_j$, where $m_j$ is the mass of an atom ($j = 0$), or an ion
with a charge $j$ ($j = 1 - 5$), or an electron ($j = e$). The
mass conservation equations of the species is then given by
\begin{equation}
 \frac{\p(\rho f_j)}{\p t} + \frac{\p}{\p r}(\rho f_j v)
=-\frac{2\rho f_j v}{r}+(S_s)_j. \label{NSr4}
\end{equation}
Here, the extra term $(S_s)_j$ is the source term for $\rho f_j$
which arises from ionization and recombination processes. It
depends on the net rate of change of the number density of the
species $\dot{n}_j$ through:
\begin{equation}
(S_s)_j=m_j\dot{n}_j.
\end{equation}
For the ions ($j=0,1,2,3,4,5$), the net rate of change is given
by:
\begin{eqnarray}
\dot{n_j} &=&  n_{j-1} n_e \alpha^{\rm ion}_{j-1\rightarrow j}
   - n_j n_e \alpha^{\rm ion}_{j\rightarrow j+1} \nonumber \\
   &&+ n_{j+1}n_e(\alpha^{\rm rrec}_{j+1\rightarrow j}
   + \alpha^{\rm trec}_{j+1\rightarrow j})
\nonumber \\
&&- n_j n_e(\alpha^{\rm rrec}_{j\rightarrow j-1}
   + \alpha^{\rm trec}_{j\rightarrow j-1}),
\label{net_rate}
\end{eqnarray}
where $\alpha^{\rm ion}_{j\rightarrow j+1}, \alpha^{\rm
rrec}_{j\rightarrow j-1}$ and $\alpha^{\rm trec}_{j\rightarrow
j-1}$ are the rates of ionization, radiative recombination and
three-body recombination of particles with a charge of $j$,
respectively. The formulas for these rates can be found in
\cite{xu2}.

The net rate of change of the number density of electrons is
simply given by charge conservation. Now that
\begin{equation}
f_e=1-\sum_{j=0}^5 f_j,
\end{equation}
taking time derivative and multiplying both sides by $\rho$ gives:
\begin{equation}
(S_s)_e = -\sum_{j=0}^5(S_s)_j.
\end{equation}

\subsection{Equation of State of gas}

The hydrodynamics of the bubble is certainly affected by the EOS.
The most widely-used van der Waals EOS can be modified to take
into account the ionization processes \cite{xu}:
\begin{equation}
P=\left(\sum_{j=0}^5\frac{f_j}{m_j}+\frac{f_e}{m_e}\right)
\frac{k_B \rho T}{1-b \rho}, \label{MVEOSp}
\end{equation}
\begin{equation}
e=\frac{3}{2}k_B T
\left(\sum_{j=0}^5\frac{f_j}{m_j}+\frac{f_e}{m_e}\right)
+k_B\sum_{j=1}^5\sum_{i=j}^5\frac{f_i}{m_i}T_j, \label{MVEOSe}
\end{equation}
where $T_j$ is the ionization energy of an ion with charge $j-1$,
and $b$ the excluded volume. This EOS is denoted as MVEOS.

The physical meanings of the MVEOS, Eqs.~(\ref{MVEOSp}) and
(\ref{MVEOSe}), are manifest. The total pressure $P$ is the sum of
the contributions by different species, which are separately taken
into account in proportion to their abundances. The internal
energy $e$ of the gas consists of both the thermal energy (the
first term) and ionization energy (the second term).

\subsection{Energy transport in the liquid}

The changes in the liquid temperature $T_l$ is accounted for with
the assumption that the liquid compressibility and viscosity do
not affect the heat transfer process between the bubble and the
surrounding water. As such, the energy equation for the water is:
\begin{equation}
\frac{\partial T_l}{\partial t}+v_l\frac{\partial T_l}{\partial r}
=D_l \frac{1}{r^2}\frac{
\partial}{\partial r}\left(r^2\frac{\partial T_l}{\partial r}
\right)\;, \label{diffeq}
\end{equation}
where $v_l$ and $D_l$ are the velocity and thermal diffusion
coefficient of the liquid, respectively.

\section{Numerical Results}\label{UPS}
In order to elucidate the significance of individual physical
factors affecting SBSL, in the following we present and compare
numerical results obtained from simulations constructed with
different models. Specifically, we consider (i) UBM versus CFM
model; (ii) GOM versus WOM; and (iii) the two plasma models, P1
versus P2.

The rest of this section is organized as follows: first we study how
the effects of various physical entities, including plasma, wave and
temperature, can affect the emitted light pulse using the simple UBM
model. Then we use the CFM model that is more realistic to mimic
SBSL and compare relevant numerical results with those of UBM.

\subsection{Emission in Uniform Bubble Model} 
\begin{figure} 
\includegraphics[width=5.7cm,angle=270]{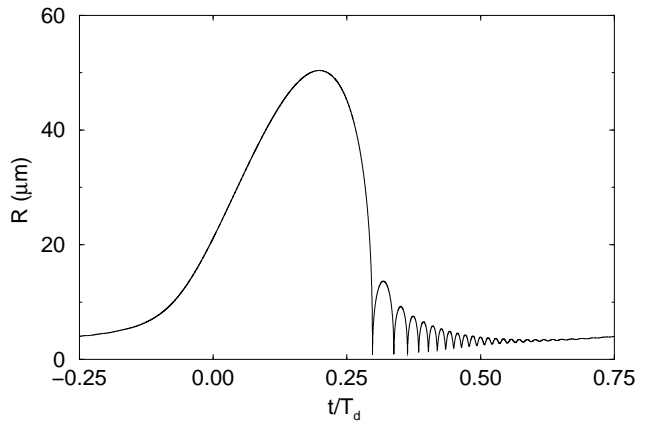}
\caption{\small A plot of the bubble radius $R$ versus time $t$,
which is normalized with respect to the acoustic period $T_{\rm
d}$.} \label{upsf1}
\end{figure}

\begin{figure}
\includegraphics[width=5cm,angle=270]{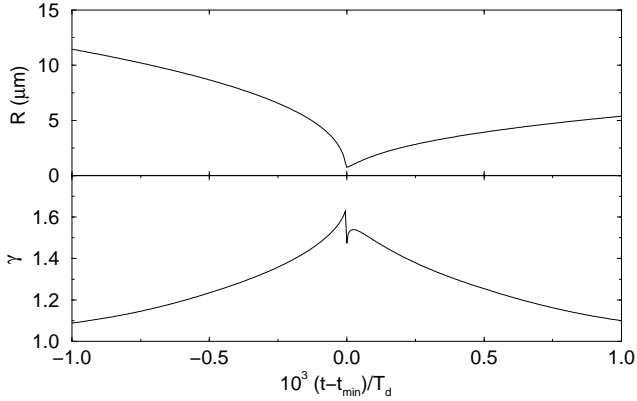}
\vspace{4mm} \caption{\small The bubble radius $R$ (upper panel)
and the polytropic exponent $\gamma$ (lower panel) are plotted
against the normalized time $10^3(t-t_{\rm min})/T_d$, where
$t_{\rm min}$ is the moment at which $R$ attains its minimum value
of about $ 0.7\mu{\rm m}$.} \label{upsf2}
\end{figure}

\begin{figure}
\includegraphics[width=5.3cm,angle=270]{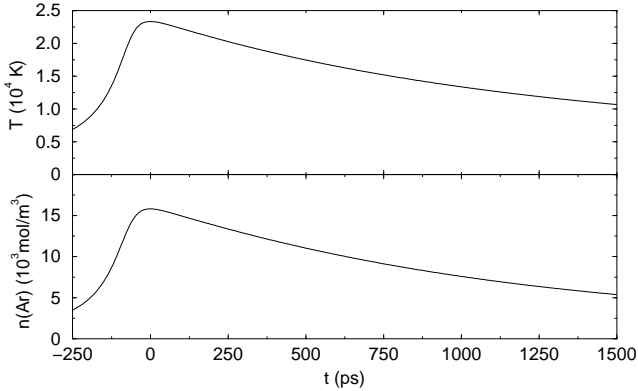}
\caption{\small Time profiles of temperature $T$ (upper panel) and
number density of argon atom $n_0$ (lower panel) near the instant
of minimum bubble size.} \label{upsf3}
\end{figure}

UBM here refers to the model used in
\cite{Lohse:1,Lohse:6}, where the RP equation assumed
incompressibility of the surrounding liquid and a variable
  polytropic exponent $\gamma(R,\dot{R},T)$ was used to account for effects of
 thermal conduction. However, instead of using the
fitting formula in \cite{Lohse:1} for computing the polytropic
exponent $\gamma(R,\dot{R},T)$, we employed the formula proposed
in \cite{Prosp:1} in the simulations. As in Ref.~\cite{Lohse:1},
we studied the oscillations of a bubble with ambient radius $R_0 =
5.0 \,\mu{\rm m}$, subjected to an ultrasonic wave with $f = 20
\,{\rm kHz}$ and $P_a = 1.3 \,{\rm atm}$. Figs.~\ref{upsf1},
\ref{upsf2} respectively show for the UBM the time evolutions of
the bubble radius $R$ over one acoustic cycle and, in the vicinity
of the maximum bubble compression, the radius and the polytropic
exponent $\gamma$.
The number density of argon atoms and the temperature near the
instant of minimum radius are shown in Fig.~\ref{upsf3}. Here we
remark that the temperature profile is slightly different from
that in \cite{Lohse:1} due to the difference in the formulas for
$\gamma(R,\dot{R},T)$. The profile remains essentially the same,
but the peak temperature in our result is about $3000 \,{\rm K}$
higher.
The temperature and density profiles are used as inputs in our
calculations of the spectral radiance, from which other light
emission properties (e.g. pulse shapes and FWHM) are obtained.

\subsubsection{Plasma and wave effects}
\label{nPS} 
\begin{figure}
\includegraphics[width=5.6cm,angle=270]{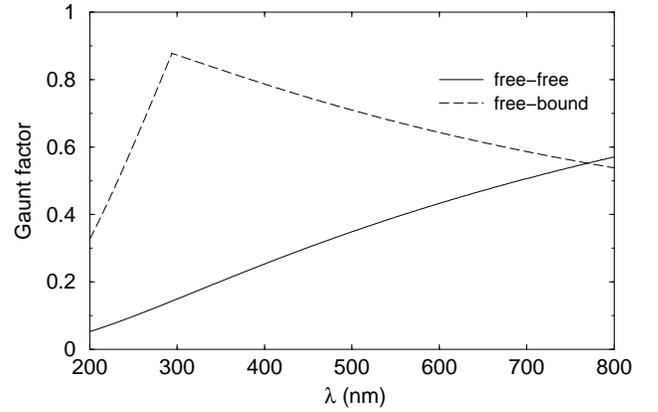}
\caption{\small Free-free (solid line) and free-bound (dashed
line) Gaunt factors multiplied by the Cillie exponential factor at
the instant of minimum bubble size. } \label{upsf7}
\end{figure}

\begin{figure}
\includegraphics[width=5.5cm,angle=270]{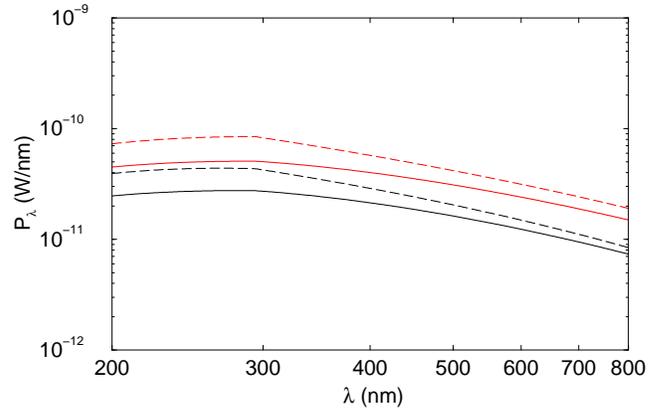}
\caption{\small (Color online) The spectral radiance $P_{\lambda}$
is plotted against the wavelength $\lambda$ for the following
models: GOM+P1 (grey-dashed line), GOM+P2 (grey-solid line),
WOM+P1 (dark-dashed line), and WOM+P2 (dark-solid line)}
\label{upsf8}
\end{figure}

As mentioned previously, we employed two plasma
 collision models P1 and P2 in the simulations. The differences in
 these two models originate from
 the free-free (or free-bound) Gaunt factor and the Cillie
exponential cut-off factor $\exp(-hf/k_BT)$. The exponential
factor  is usually close to unity in the Rayleigh-Jeans limit
where $hf \ll k_BT$, but for the case of SL, $k_BT$ is of the
order of a few eV and is within the range $1.5\,-\,6.2\,\rm{eV}$
of the observed light spectrum, and hence is not negligible. In
particular we find, when the bubble is at minimum size and the
temperature and density of its contents are also at their maxima,
the free-free Gaunt factor and the free-bound Gaunt factors,
multiplied by the exponential cut-off, result in a
 correction factor of order $0.1$ (see Fig.~\ref{upsf7}).
Fig.~\ref{upsf8} shows the computed power spectra obtained from
simulations constructed respectively with GOM/WOM+P1/P2, clearly
demonstrating that the P2 model indeed leads to an decrease in the
radiance.

Furthermore, we can observe in Fig.~\ref{upsf8} that the power is
overestimated when GOM is used rather than WOM, which can be
readily explained as follows. When a plane wave is incident on a
WOM bubble, part of it is reflected or scattered from the
boundaries and the remaining part is absorbed as heat. In light of
Kirchhoff's law then, less absorption implies less emission. In
contrast, reflection and diffraction are neglected in GOM,
resulting in an overestimated absorption and hence emission.
Therefore, to achieve realistic power calculations comparable with
experimental results, the effects of both the wave nature of light
and the Gaunt factor correction cannot be neglected, consistent
with the point we made earlier.
%
%

In subsequent discussions we will employ two specific light
emission models: the model proposed in Ref.~\cite{Lohse:1}
(GOM+P1) and our present model (WOM+P2), and it should be
understood that all power computations employing GOM are done with
P1 while those employing WOM are done with P2.
\subsubsection{Temperature effects} \label{intemp} 
\begin{figure}
\includegraphics[width=5.5cm,angle=270]{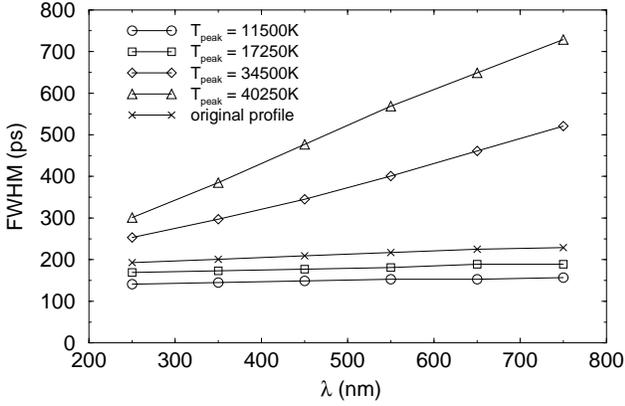}
\caption{\small FWHM obtained from UBM+GOM versus wavelength
$\lambda$. Lines with circles, squares, crosses, diamonds and
triangles respectively represent the cases with a scaling of
$0.5$; $0.75$; 1 (i.e. no scaling); $1.50$; and $1.75$.}
\label{upsf14}
\end{figure}

\begin{figure}
\includegraphics[width=5.5cm,angle=270]{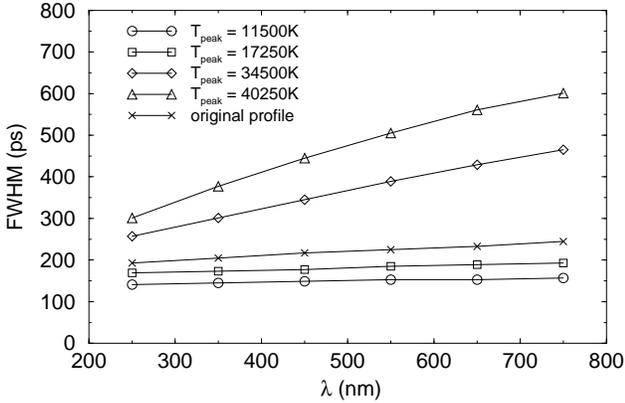}
\caption{\small FWHM obtained from UBM+WOM versus wavelength
$\lambda$. Lines with circles, squares, crosses, diamonds and
triangles respectively represent the cases with a scaling of
$0.5$; $0.75$; 1 (i.e. no scaling); $1.50$; and $1.75$.}
\label{upsf15}
\end{figure}

\indent For our
case studied here using the UBM model, the maximum temperature
achieved at the instant of minimum bubble size is about $23000
\,{\rm K}$ (Fig.~\ref{upsf3}). To study the effect of the interior
temperature on light emission, we scale by hand the original
temperature profile for $P_a = 1.300 \, {\rm atm}$ and $R_0 = 5.0
\,\mu{\rm m}$ by some chosen factors, say $0.3, 0.5, 0.75, 1.5$
and $1.75$, while keeping the densities of Ar neutrals constant.
Accordingly, two of the three input parameters
 to the light emission model (temperature and ion number density) are
changed and one (the atom number density) remains fixed. We employ
both GOM and WOM to calculate the power, and study the spectral
variation of the FWHM calculated within $100\,{\rm nm}$ wavelength
windows as shown in Figs.~\ref{upsf14} and \ref{upsf15}.
It is remarkable that a consistently smaller spectral variation of
the FWHM is obtained when the light emission model WOM is used.
The unscaled profile is about constant over the range of
wavelength considered as in Ref.~\cite{Lohse:1}, which employed
UBM and GOM, and the profile still remains remarkably constant
when the temperature was scaled down by a factor of $0.5$ and
$0.75$ (to a peak temperature of about $10000-20000 \,{\rm K}$).
Scaling down by $0.3$ (to a peak temperature of about $ 7000 \,
{\rm K}$) produced zero power output since the temperature was
much lower than that required for ionization. Nevertheless,
scaling up by $1.5$ and $1.75$ (to a peak temperature of about $
35000-45000 \,{\rm K}$) produces dramatic variation of the FWHM.
Hence, we find that the spectral uniformity of the FWHM holds only
when the SL bubble temperature is restricted within a rather small
range of moderate values. However, we remark that the temperature
is underestimated in the UBM since local temperature rises were
not taken account of. With a more realistic hydrodynamic
modelling, Ref.~\cite{Ho:1} found the temperature should be
several $10^4~\textrm{K}$ higher. Thus, we expect that a larger
FWHM spectral variation with increased driving pressure will be an
essential realistic feature of SL.

In addition, we have studied the effect of ambient water
temperature on SL light emission, using WOM and the values of
$(P_a,~R_0)$ extracted from the phase diagram in
Ref.~\cite{Lohse:3}, at $T_0 = 20 \,^{\rm o}{\rm C}$ and $2.5
\,^{\rm o}{\rm C}$ and at a driving frequency of $f = 26.5 \,{\rm
kHz}$.
\begin{table}[htbp]
\begin{center}
\vspace{5mm}
\begin{tabular}{|c|c|c|c|c|}
\hline
\multicolumn{5}{|c|}{$T_0 = 20\,^{\mathrm{o}}\mathrm{C}$}
\\ \hline
$P_a$ (atm) & 1.275 & 1.283 & 1.292 & 1.300
\\ \hline
$R_0$ ($\mu$m) & 2.9 & 3.2 & 3.6 & 3.9
\\ \hline
\multicolumn{5}{|c|}{$T_0 = 2.5\,^{\mathrm{o}}\mathrm{C}$}
\\ \hline
$P_a$ (atm) & 1.320 & 1.350 & 1.375 & 1.400
\\ \hline
$R_0$ ($\mu$m) & 2.0 & 3.5 & 4.0 & 4.5
\\ \hline
\end{tabular}
\caption{\small Driving pressure $P_a$ and ambient radius
$R_0$ at $T_0 = 20 \,^{\rm o}{\rm C}$ and $T_0 = 2.5 \,^{\rm o}{\rm C}$
                that give stable sonoluminescing bubble for $f=26.5 \,{\rm kHz}$.}
\label{ups_t1}
\end{center}
\end{table}
%
%
\begin{figure}
\includegraphics[width=5.5cm,angle=270]{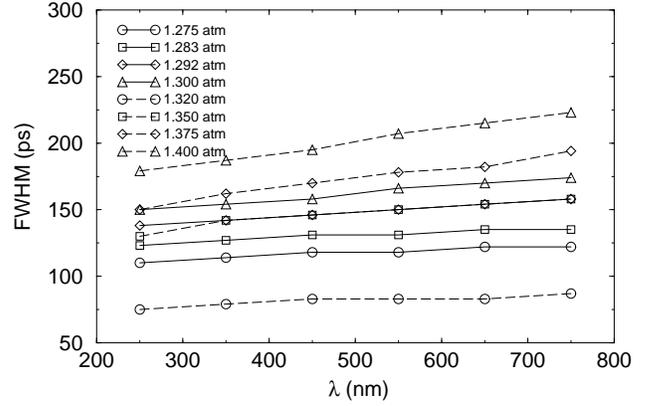}
\caption{\small FWHM versus wavelength $\lambda$ calculated from
UBM+WOM. Cases with $T = 20 \,^{\rm o}{\rm C}$ and $T = 2.5
\,^{\rm o}{\rm C}$ are indicated by the solid curve and  the
dashed curve, respectively.} \label{upsf17}
\end{figure}
Here we observe from Fig.~\ref{upsf17} the general trend of a
larger FWHM towards the red end of the spectrum as the pressure is
tuned up, and this increase is further enhanced at a lower water
temperature. This is consistent with the experimental findings of
Moran {\it et al.} \cite{pulse:1} and is readily explained. At a
lower water temperature, the bubble can be driven harder
\cite{Lohse:3} so that, given a certain value of ambient radius, a
larger driving pressure can be applied while  maintaining bubble
stability. The bubble collapses more violently under the larger
pressure and hence the temperature of the bubble interior achieves
a higher value, resulting in a larger FWHM spectral variation.
Therefore, in effect, both of our observations under increased
driving pressure and lower water temperature can be explained in
terms of the higher temperature reached inside the bubble.
\subsection{Computational Fluid Mechanics Model}\label{nhydro}
\subsubsection{Hydrodynamics}
\begin{figure}
\includegraphics[width=4.8cm,angle=270]{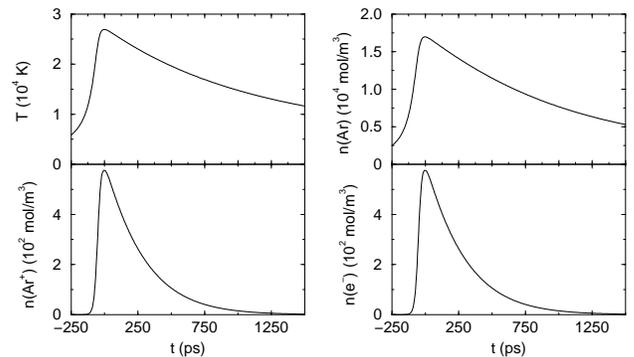}
\caption{\small Temperature $T$, number densities of argon
neutrals $n({\rm Ar})$, ions $n({\rm Ar}^+)$ and electrons $n({\rm
e}^-)$ shown as a function of time, where $t=0$ is the instant of
minimum bubble radius, for a UBM bubble with $P_a = 1.325 \,{\rm
atm}$, and $R_0 = 4.7 \,\mu{\rm m}$.} \label{upsf20}
\end{figure}

\begin{figure}
\includegraphics[width=5.3cm,angle=270]{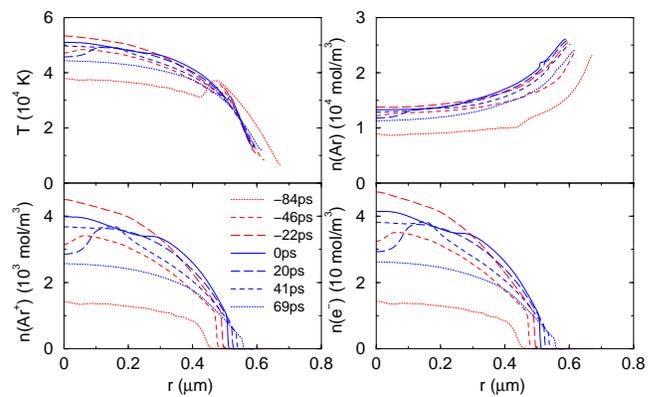}
\caption{\small (Color online)  Snapshots near the instant of
minimum bubble radius ($t=0$)
 for the same quantities as in Fig.~\ref{upsf20} plotted
against radial distance, for a CFM bubble with $P_a = 1.325 \,{\rm
atm}$, and $R_0 = 4.7 \,\mu{\rm m}$. Grey lines indicate times
before zero while dark lines indicate times after zero.}
\label{upsf28}
\end{figure}

Now we employ CFM developed by Ho {\it et al.} \cite{Ho:1} that
includes the effects of ionizations and recombinations to
calculate the power spectra and pulse profiles using WOM, and
compare the results with those obtained from joint application of
UBM and GOM. The set of conditions that we employ is extracted
from Ref.~\cite{Lohse:1} and shown in Table~\ref{ups_t3}, where
$T_0 = 20 \,^{\rm o}{\rm C}, f = 20 \,{\rm kHz}$ and the dissolved
gas concentration is $0.20\%$. Figs.~\ref{upsf20} and \ref{upsf28}
respectively show  the computed results of UBM and CFM for a case
with $P_a = 1.325 \,{\rm atm}$, $R_0 = 4.7 \,\mu{\rm m}$. It is
found that the maximum temperature obtained with CFM can exceed $5
\times 10^4 \,{\rm K}$ while that in UBM is less than $3 \times
10^4 \,{\rm K}$. Accordingly, the number of Ar$^+$ ion in CFM is
much greater than that in UBM.
\begin{table}[htbp]
\begin{center}
\vspace{5mm}
\begin{tabular}{|c|c|c|c|c|}
\hline $P_a$ (atm) & 1.275 & 1.300 & 1.325 & 1.350
\\ \hline
$R_0$ ($\mu$m) & 2.6 & 4.0 & 4.7 & 5.4
\\ \hline
\end{tabular}
\caption{\small Driving pressure $P_a$ and ambient radius $R_0$ at
$T_0 = 20 \,^{\rm o}{\rm C}$ that give stable sonoluminescing bubble
for $f=20 \, {\rm kHz}$.} \label{ups_t3}
\end{center}
\end{table}

\subsubsection{Light emission}

To apply WOM to CFM, which produces an inhomogeneous profile of
bubble temperature and number densities, we approximate the
resultant inhomogeneous profile by a layered one and use
 within each layer $j$ the average values of the temperature,
 the number densities of atom and electron  there. The absorption coefficient $\kappa(j)$,
collision frequency $\gamma(j)$ and refractive index $n(j)$ can
accordingly be computed using these averaged values. This
approximation scheme allows for the application of WOM to the
resulting multilayered spherical configuration.

\begin{figure}
\includegraphics[width=5cm,angle=270]{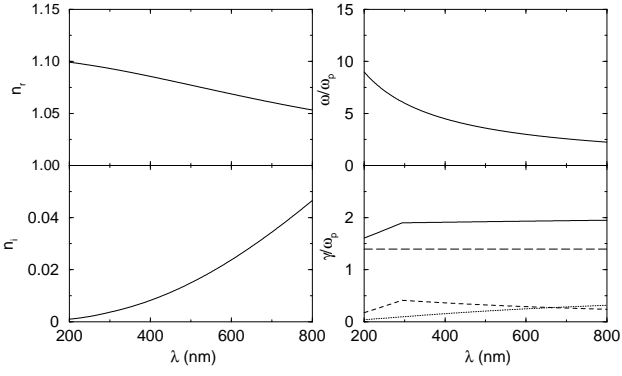}
\caption{\small Shown on the left are the real part $n_R$ and
imaginary part $n_I$ of the refractive index, on the right are the
ratios $\omega/\omega_p$ and $\nu/\omega_p$ versus the wavelength
$\lambda$ for a UBM bubble with $P_a = 1.375 \,{\rm atm}$ and $R_0
= 2.6\,\mu{\rm m}$ at the instant of minimum radius. In the graph
$\nu/\omega_p$ versus $\lambda$, the dotted, short-dashed,
long-dashed and full lines respectively show $\nu_{ei}$,
$\nu_{rc}$, $\nu_{ea}$ and $\nu$. } \label{upsf32}
\end{figure}

\begin{figure}
\includegraphics[width=5cm,angle=270]{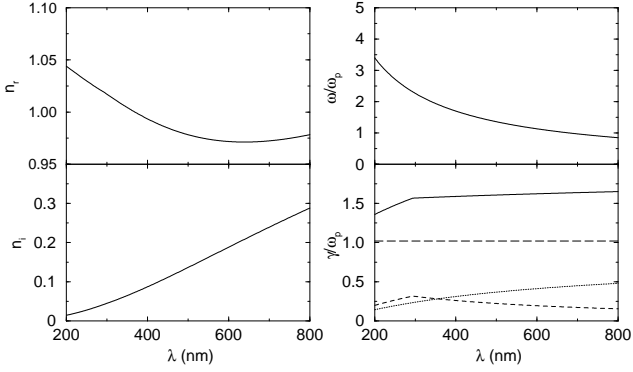}
\vspace{0mm} \caption{\small Same as Fig.~\ref{upsf32}, for a CFM
bubble with $P_a = 1.325 \,{\rm atm}$, and $R_0 = 4.7 \,\mu{\rm
m}$. The quantities shown here are those of the innermost layer
which occupies the inner $7.5\%$ of the bubble radius for $P_a =
1.325 \,{\rm atm}$ hence is representative of the hottest and
densest region of the bubble.} \label{upsf36}
\end{figure}

As the degree of ionization in CFM result is much higher than that
in UBM, the difference in the optical properties of these two
models is obvious. As shown in Figs.~\ref{upsf32} and
\ref{upsf36}, $\omega/\omega_p$ is reduced by a factor of $2.5$ in
CFM as compared to UBM. In particular, in CFM $\omega/\omega_p$ is
close to unity near the UV end, indicating that plasma collective
effects may be significant in the short-wavelength regime. Also,
both dispersion and absorption are considerably stronger in the
CFM case, exhibiting a larger variation in $n_R$ and a larger
$n_I$ than the UBM case.

\begin{figure}
\includegraphics[width=5cm,angle=270]{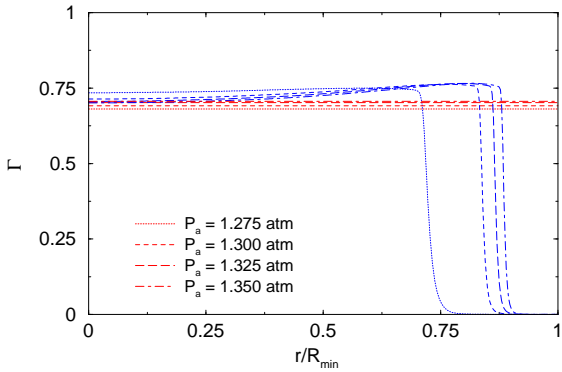}
\caption{\small (Color online) The ion-coupling parameter $\Gamma$
is plotted against $r$, normalized by the minimum radius $R_{\rm
min}$ at the instant of maximal bubble compression.  The values
obtained from UBM (CFM) are indicated by grey (dark) lines. }
\label{upsf41}
\end{figure}

In Sec.~\ref{PS} we have made the assumption that the plasma is so
tenuous that the Coulomb energy is much smaller than the average
thermal energy of individual particles and the plasma behaves like
an ideal gas. The values of the ion-coupling parameter $\Gamma$ at
the instant of maximum bubble compression for the cases studied
here (see Table~\ref{ups_t3}) are shown in Fig.~\ref{upsf41}.
For UBM we note that $\Gamma \simeq 0.7$ for all four cases. In
the CFM bubble, interestingly there is a clear formation of two
regions: an inner core which is moderately coupled $\Gamma \sim
0.7$ and an outer shell which is weakly coupled $\Gamma \simeq 0$.
In addition the effect of increasing the driving pressure is seen
to increase the size of this moderately-coupled inner core. Since
the degree of coupling is moderate and not too strong, we expect
that the formulas used for the absorption coefficients and
collision frequencies based on the tenuous plasma assumption
should still apply. However, on the other hand, if the driving
pressure increases while maintaining the stability of oscillation,
it is likely that the plasma might become a non-ideal one. The
physical property of such dense non-ideal plasma is rather
complicated and is beyond the scope of the present paper
\cite{Fortov:1}.

In the following we contrast data obtained respectively from
GOM+UBM and WOM+CFM and specifically consider three different
physical quantities, namely the spectrum, the pulse shape and the
FWHM of light pulses.

\begin{figure}
\includegraphics[width=5cm,angle=270]{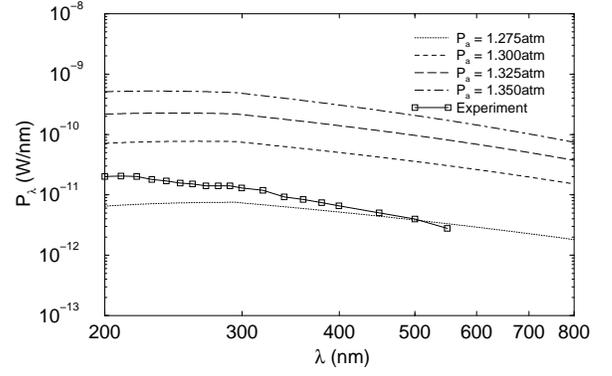}
\vspace{0mm}\caption{\small Spectral radiance $P_{\lambda}$ versus
wavelength $\lambda$ obtained from GOM+UBM+P1. The experimental
spectrum is obtained from Barber {\it et al.} \cite{Barber:1}}
\label{upsf22}
\end{figure}

\begin{figure}
\includegraphics[width=5.6cm,angle=270]{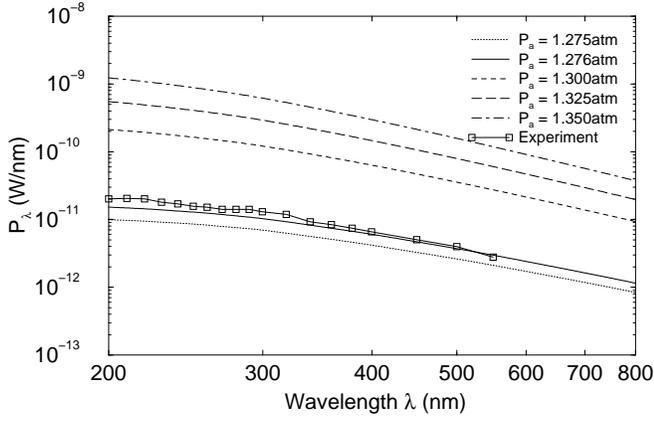}
\caption{\small Spectral radiance $P_{\lambda}$ versus wavelength
$\lambda$ obtained from WOM+CFM+P2. The best-fit for the
experimental spectrum \cite{Barber:1} is indicated by the solid
line, in which $P_a = 1.276\,{\rm atm}$, $R_0 = 2.7\,\mu{\rm m}$.}
\label{upsf38}
\end{figure}

In Figs.~\ref{upsf22} and \ref{upsf38} we show the computed
spectra using GOM+UBM+P1 and WOM+CFM+P2, respectively. One clearly
sees the improvement (Fig.~\ref{upsf38}) of our refined model,
namely WOM+CFM+P2, that the calculated spectral shape is much
closer to the experimental results \cite{Barber:1} than the
GOM+UBM+P1 model. In particular, major improvement is seen in the
UV portion of the spectrum.

\begin{figure}
\includegraphics[width=5.6cm,angle=270]{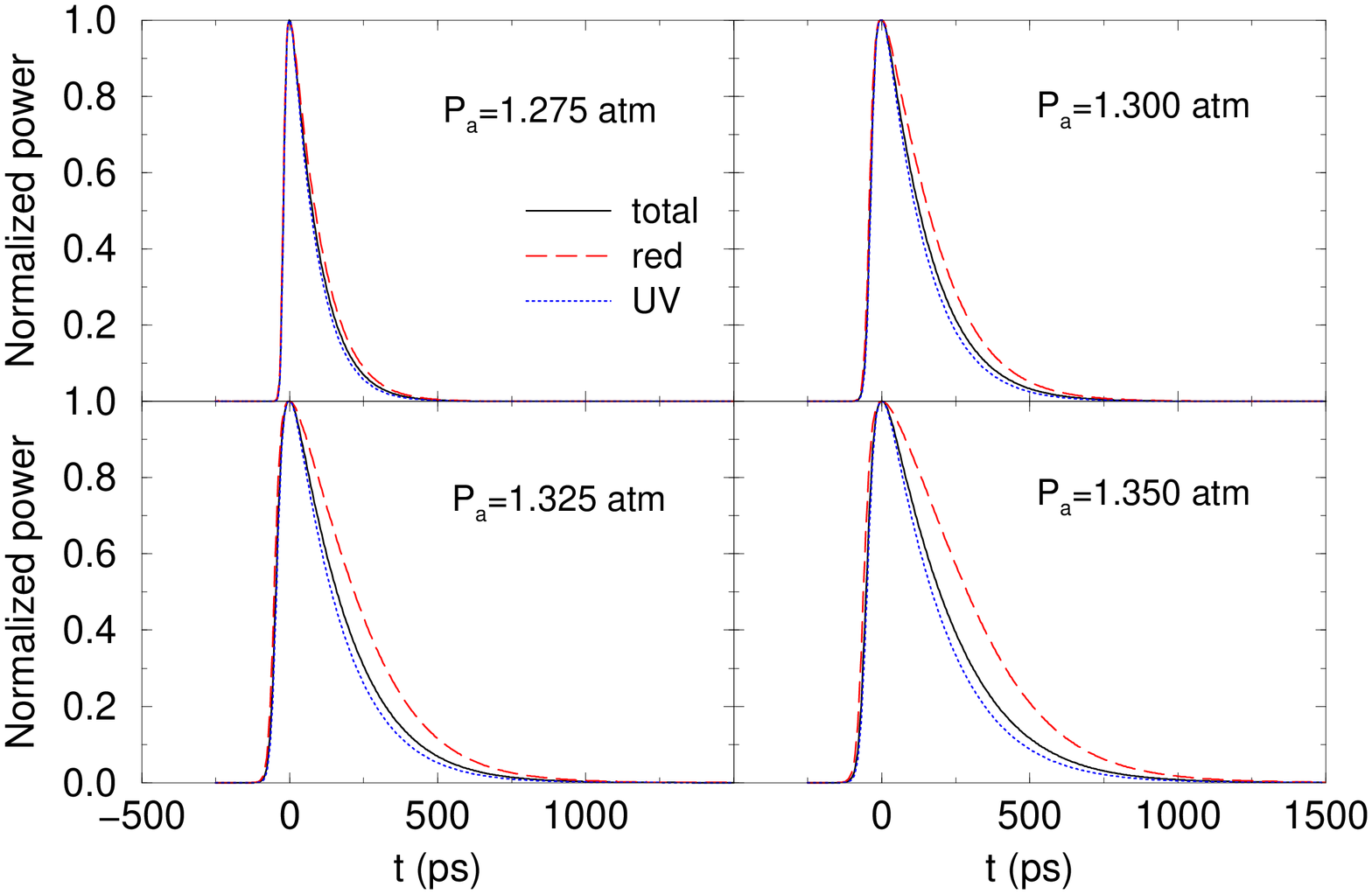}
\caption{\small (Color online)  Normalized power versus time
obtained from GOM+UBM+P1. The solid, dashed, and dotted lines are
 respectively the total normalized power, the normalized powers in the red and
UV regions.} \label{upsf24}
\end{figure}

\begin{figure}
\includegraphics[width=5.6cm,angle=270]{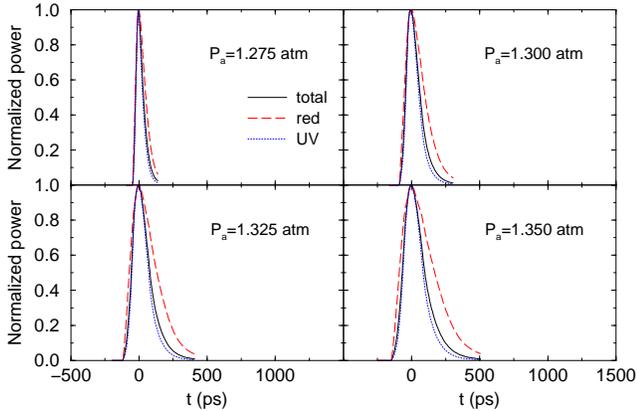}
\vspace{0mm} \caption{\small (Color online)  Normalized power
versus time obtained from WOM+CFM+P2. The solid, dashed, and
dotted lines are
 respectively the total normalized power, the normalized powers in the red and
UV regions.} \label{upsf39}
\end{figure}

The calculated pulse shapes are shown in Figs.~\ref{upsf24} and
~\ref{upsf39} respectively, evidently the pulse shapes produced
from WOM+CFM+P2 are more consistent with experimental data
\cite{pulse:1,pulse:2,pulse:4} where the long-time tail was not
observed. By contrast, as shown Fig.~\ref{upsf24}, a long-time
tail appears in GOM+UBM+P1 and is an undesirable feature.

\begin{figure}
\includegraphics[width=5.6cm,angle=270]{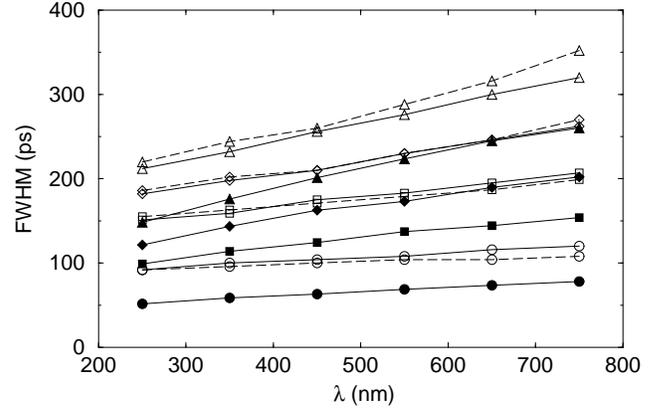}
\caption{\small FWHM of light pulse versus wavelength $\lambda$
obtained at different driving pressures: circle $P_a = 1.275\,{\rm
atm}$, square $P_a = 1.300\,{\rm atm}$, diamond $P_a = 1.325\,{\rm
atm}$ and triangle $P_a = 1.350\,{\rm atm}$. Dashed lines with
empty symbols and solid lines with filled symbols  are calculated
from GOM+UBM+P1 and WOM+CFM+P2, respectively.} \label{upsf40}
\end{figure}

Fig.~\ref{upsf40} shows the calculated FWHM plotted against the
wavelength for both cases. While it is clear that the FWHM increases
with driving pressure as found in experiments
\cite{pulse:1,pulse:2,pulse:4}, we also see that the FWHM remains
nearly a constant over $200\,-\,800\,\rm{nm}$ \textit{only} at a low
pressure $P_a \sim 1.275\,{\rm atm}$ even in the more realistic
WOM+CFM+P2 model (c.f. Sec.~\ref{intemp}). We remark that the
results obtained by Gompf {\textit{et al}.} \cite{pulse:4} showing
similar pulse widths for the red and UV pulse were obtained under a
driving pressure of $P_a = 1.200\,{\rm atm}$; which is smaller than
the lowest pressure $P_a = 1.275\,{\rm atm}$ we used and is
therefore expected to show a constant pulse width. In particular the
results of Moran \textit{et al}. \cite{pulse:1} showed the FWHM
increases with wavelength at a low ambient water temperature
$3\,^{\rm{o}}\rm{C}$, which, as remarked previously,  also resulted
in a larger driving pressure. Thus we emphasize that in general the
notion of wavelength-independence of the SL pulse width is only
correct at low driving pressures; and at higher driving pressures
spectral dispersion of the pulse width shows up, and this can be
simply explained within our model. Either effects of higher driving
pressure or lower water temperature boil down to the consequence of
higher bubble temperature. As the bubble becomes hotter, both
optical dispersion and absorption become more significant and hence
the bubble becomes more optically opaque, approaching a blackbody
surface emitter. Red light is then radiated for a longer duration
than the UV since, throughout one cycle, the bubble can stay at a
lower-temperature state for a longer duration. As a consequence, the
pulse width increases towards the red end of the spectrum. In other
words, the emitted light becomes more spectrally dispersive because
of the increased absorption and dispersion in the plasma medium. In
particular, the absorption is highest (Figs.~\ref{upsf32} and
\ref{upsf36}) at the red end of the spectrum, resulting in a flatter
pulse shape there (larger FWHM). Thus,
 besides using a lower ambient temperature as in Ref.~\cite{pulse:1},
if a strong enough pressure is applied (while still maintaining
bubble stability) the spectral variation of the FWHM would be an
observable consequence in experiment.
\section{Conclusion}\label{sum}
In summary, we have proposed in the present paper a robust theory
for optical emission in SBSL that properly takes into account of
the wave nature and propagation of light in the absorptive plasma
formed inside a sonoluminescing bubble in a self-consistent way.
In addition, our theory can be applied to bubbles with
inhomogeneous density and temperature profiles. The validity of
our scheme was examined for the case of SBSL with argon bubbles.
By introducing and implementing appropriate Gaunt factors and
exponential correction in the collision frequencies; as well as
the effects of optical thickness, scattering, reflection and
diffraction, our light emission model successfully explains the
major features (including power spectrum, pulse shape and FWHM)
observed in SBSL experiments. In addition, the computed power
spectra and pulse shapes are shown to be in excellent agreement
with experimental results.

Besides, we have also shown that the experimentally observed
spectral independence of the FWHM at $T_0 = 20 \,^{\rm o}{\rm C}$
is ascribable to the relatively small temperature range (about
$10000-30000 \,{\rm K}$) achievable in a collapsing SL bubble.
Above this range the bubble behaves in the way of a blackbody
surface emitter and the spectral variation of the FWHM should be
more notable. In fact, as the driving pressure goes up, the
temperature reached inside the bubble rises. Also, if the ambient
water temperature is lowered at a fixed ambient radius $R_0$, the
driving pressure allowable for bubble stability extends to a
larger value. This provides a theoretical explanation for why
Moran {\it et al.} \cite{pulse:1} found a spectral variation of
the pulse width at $T = 3 \,^{\rm o}{\rm C}$.

Notwithstanding the above-mentioned achievements, the model
developed in the present paper is only one of the many steps towards
a better understanding of SBSL, which is an extremely complex
phenomenon resulting from the subtle interplay of hydrodynamics,
chemical reactions, plasma physics and optics as well. Much more
challenging problems, e.g. SBSL with inert gases other than argon
and inclusion of water vapor in the hydrodynamic code, are still
ahead for us. They are surely our goal of endeavor in the future.
\acknowledgments We thank M.-C. Chu and K.M. Pang for discussions.
The work described in this paper was partially supported by a
grant (Project No. 401603) from the Research Grants Council of the
Hong Kong Special Administrative Region, China.

\newcommand{\noopsort}[1]{} \newcommand{\printfirst}[2]{#1}
  \newcommand{\singleletter}[1]{#1} \newcommand{\switchargs}[2]{#2#1}

\end{document}